\documentclass[conference]{IEEEtran}
\IEEEoverridecommandlockouts
\usepackage{booktabs} 
\usepackage{cite}
\usepackage{amsmath,amssymb,amsfonts}
\usepackage{url}
\usepackage{hyperref}
\usepackage{textcomp}
\usepackage{xcolor}
\usepackage{enumitem}
\usepackage{extarrows} 
\usepackage{graphicx}
\usepackage{subcaption}
\usepackage[para,online,flushleft]{threeparttable}
\usepackage{tabularx}
\usepackage{amsmath}
\usepackage{stmaryrd}
\usepackage{makecell}
\usepackage{caption}
\usepackage{color, colortbl}
\usepackage[ruled,vlined,noend]{algorithm2e}
\usepackage{algpseudocode}
\usepackage{adjustbox}
\usepackage{pgfplots}
\usepgfplotslibrary{statistics}
\usepackage{soul}
\usepgfplotslibrary{fillbetween}
\usepackage{filecontents}
\usepackage{balance}
\usepackage{booktabs}
\usetikzlibrary{patterns}
\usepackage{multirow}
\pgfplotsset{width=8cm,compat=1.9, max space between ticks=35}
\newcommand*\circled[1]{\tikz[baseline=(char.base)]{
            \node[shape=circle,draw,inner sep=1pt] (char) {#1};}}
\renewcommand{\thetable}{\arabic{table}}
\pgfplotsset{width=8cm,compat=1.9, max space between ticks=35}
\def\BibTeX{{\rm B\kern-.05em{\sc i\kern-.025em b}\kern-.08em
    T\kern-.1667em\lower.7ex\hbox{E}\kern-.125emX}}
\usetikzlibrary{shapes.geometric}

\begin{document}

\title{\textsc{Metasql}: A Generate-then-Rank Framework for Natural Language to SQL Translation}

\author{\IEEEauthorblockN{Yuankai Fan, Zhenying He, Tonghui Ren, Can Huang, Yinan Jing, Kai Zhang, X.Sean Wang}
\IEEEauthorblockA{\textit{School of Computer Science}, \textit{Fudan University}, Shanghai, China \\
fanyuankai@fudan.edu.cn, zhenying@fudan.edu.cn, thren22@m.fudan.edu.cn \\
huangcan22@m.fudan.edu.cn, jingyn@fudan.edu.cn, zhangk@fudan.edu.cn, xywangCS@fudan.edu.cn}
}

\maketitle

\begin{abstract}
The Natural Language Interface to Databases (NLIDB) empowers non-technical users with database access through intuitive natural language (NL) interactions. Advanced approaches, utilizing neural sequence-to-sequence models or large-scale language models, typically employ auto-regressive decoding to generate unique SQL queries sequentially. While these translation models have greatly improved the overall translation accuracy, surpassing 70\% on NLIDB benchmarks, the use of auto-regressive decoding to generate single SQL queries may result in sub-optimal outputs, potentially leading to erroneous translations. In this paper, we propose \textsc{Metasql}, a unified \textit{generate-then-rank} framework that can be flexibly incorporated with existing NLIDBs to consistently improve their translation accuracy. \textsc{Metasql} introduces query metadata to control the generation of better SQL query candidates and uses learning-to-rank algorithms to retrieve globally optimized queries. Specifically, \textsc{Metasql} first breaks down the meaning of the given NL query into a set of possible query metadata, representing the basic concepts of the semantics. These metadata are then used as language constraints to steer the underlying translation model toward generating a set of candidate SQL queries. Finally, \textsc{Metasql} ranks the candidates to identify the best matching one for the given NL query. Extensive experiments are performed to study \textsc{Metasql} on two public NLIDB benchmarks. The results show that the performance of the translation models can be effectively improved using \textsc{Metasql}. In particular, applying \textsc{Metasql} to the published \textsc{Lgesql} model obtains a translation accuracy of 77.4\% on the validation set and 72.3\% on the test set of the \textsc{Spider} benchmark, outperforming the baseline by 2.3\% and 0.3\%, respectively. Moreover, applying \textsc{Metasql} to \textsc{Gpt-4} achieves translation accuracies of 68.6\%, 42.0\%, and 17.6\% on the three real-world complex scientific databases of \textsc{ScienceBenchmark}, respectively. The code for \textsc{Metasql} is available at \url{https://github.com/Kaimary/MetaSQL}.


\end{abstract}

\begin{IEEEkeywords}
NLIDB, NL2SQL, SQL, learning-to-rank
\end{IEEEkeywords}

\definecolor{pp}{rgb}{0.53, 0, 0.50}
\definecolor{rd}{rgb}{0.77, 0, 0}
\section{Introduction}
 Making databases accessible is as important as the performance and functionality of databases. Many techniques, such as natural language (NL) interfaces, have been developed to enhance the ease of use of databases in the last few decades \cite{Androutsopoulos95, Visionnary99, Banks02}. These NL interfaces democratize database access for ordinary users who may not be proficient in query languages (e.g., SQL). As a result, the construction of natural language interfaces to databases (NLIDB) has garnered extensive attention from the data management and natural language processing (NLP) communities.

 Prior works have explored machine-learning methods that employ either neural sequence-to-sequence (Seq2seq) models \cite{sqlnet17, IRNet19, Bogin19, BoginGB19, RATSQL20, GAP21, SmBop20, LGESQL20, resdsql23} or large-scale language models (LLMs) \cite{GPT4, DINSQL23, SQL-PaLM23} to generate distinct SQL outputs via auto-regressive decoding. However, despite achieving notable gains in translation accuracy, unsatisfactory performance of these approaches was observed in the overall improvement. For example, the state-of-the-art model on top of the widely used \textsc{Spider} \cite{Yu18} benchmark\footnote{Refer to the leaderboard \url{https://yale-lily.github.io/spider}. Note that \textsc{Spider} uses unknown testing queries and databases to evaluate NLIDB algorithms.} attains only 74.0\% accuracy in syntactic equivalence translation on the test set at the time of writing\footnote{\textsc{Spider} provides two separate leaderboards to assess NLIDB algorithms, one for syntactic equivalence accuracy and the other for execution equivalence accuracy. This paper primarily focuses on the former, given that most existing Seq2seq-based translation models do not predict specific values in SQL queries, which makes the latter evaluation unsuitable for them.}.

 

\newcommand\boldpp[1]{\textcolor{pp}{\textbf{#1}}}
\definecolor{darkred}{rgb}{0.70, 0, 0}
\definecolor{pale}{rgb}{0.98, 0.91, 0.85}
\begin{figure}
 \begin{subfigure}[b]{1.0\linewidth}
 \begin{adjustbox}{width=.46\linewidth}
    \begin{tabular}{c c c c}
        \hline
        \rowcolor{pale} \textbf{countryCode} & \textbf{language} & \textbf{isOfficial} & \textbf{percentage} \\ \hline
        ABW & Dutch & T & 5.3 \\ 
        ABW & English & F & 9.5 \\ 
        ABW & Papiamento & F & 76.7   \\ 
        ABW & Spanish & F & 7.4   \\ 
        AFG & Balochi & F & 0.9 \\
        AFG & Dari & T & 32.1 \\
        AFG & Pashto & T & 52.4  \\ 
        AFG & Turkmenian & F & 1.9\\
        AFG & Uzbek & F & 8.8 \\ 
        BMU & English & T & 100.0 \\ 
        ... \\ \hline 
    \end{tabular}
\end{adjustbox} \hspace{2mm}
 \begin{adjustbox}{width=.46\linewidth}
    \begin{tabular}{c c c c}
        \hline
        \rowcolor{pale} \textbf{code} & \textbf{name} & \textbf{continent} & \textbf{population} \\ \hline
        ABW & Aruba & North America & 103000 \\ 
        AFG & Afghanistan & Asia & 22720000 \\ 
        AIA & Anguilla & North America & 8000   \\ 
        BMU & Bermuda & North America & 65000 \\
        CHE & Switzerland & Europe & 7160400  \\
        CMR & Cameroon & Africa & 15085000 \\ 
        COL & Columbia & South America & 42321000 \\ 
        GEO & Georgia & Asia & 4968000 \\ 
        GRC & Greece & Europe & 10545700 \\
        ISR & Israel & Asia & 6217000 \\
        ...\\ \hline 
    \end{tabular}
    \end{adjustbox}
 \vspace{-1mm}
 \captionsetup{font=small}
 \caption{A simplify database: \footnotesize{\textit{CountryLanguage} (left)} and \footnotesize{\textit{Country} (right)} tables.}
 \label{fig:(a)spider query}
 \end{subfigure}
 
 \begin{subfigure}[b]{1.0\linewidth}
  \centering
  \begin{adjustbox}{width=\textwidth}
  \begin{tabular}{l l} \Xhline{3\arrayrulewidth}
     NL \;\;Query: & \emph{\makecell[l]{What are the country codes for countries that do not speak English?}}  \\
     \makecell[l]{SQL \textbf{(Gold)}:} &
     \makecell[l]{{\fontfamily{pcr}\selectfont \textcolor{pp}{\textbf{SELECT}} countrycode \textcolor{pp}{\textbf{FROM}} CountryLanguage \textcolor{pp}{\textbf{EXCEPT}}}
     \\ {\fontfamily{pcr}\selectfont \textcolor{pp}{\textbf{SELECT}} countrycode \textcolor{pp}{\textbf{FROM}} CountryLanguage \textcolor{pp}{\textbf{WHERE}} language=\textcolor{rd}{\textquotesingle English\textquotesingle}}}
     \\[-2.0ex] \\ \hline \hline & \\[-2.0ex]
     & \makecell[c]{Beam search outputs from \textsc{Lgesql} model \cite{LGESQL20}} \\
     \makecell[l]{Top-1 SQL:} & \makecell[l]{{\fontfamily{pcr}\selectfont \boldpp{SELECT} country\textbf{code} \boldpp{FROM} \textbf{CountryLanguage} \boldpp{WHERE} \textbf{language}!=\textbf{\textquotesingle value\textquotesingle}}}  \\
     \makecell[l]{Top-2 SQL:} & \makecell[l]{{\fontfamily{pcr}\selectfont \boldpp{SELECT} \textbf{code} \boldpp{FROM} \textbf{CountryLanguage} \textcolor{pp}{JOIN} Country \boldpp{WHERE} \textbf{language}!=\textbf{\textquotesingle value\textquotesingle}}} \\
     Top-3 SQL: & \makecell[l]{{\fontfamily{pcr}\selectfont \boldpp{SELECT} country\textbf{code} \boldpp{FROM} \textbf{CountryLanguage} \boldpp{WHERE} \textbf{language}<=\textbf{\textquotesingle value\textquotesingle}}}  \\
     Top-4 SQL: & \makecell[l]{{\fontfamily{pcr}\selectfont \boldpp{SELECT} \textbf{code} \boldpp{FROM} \textbf{CountryLanguage} \textcolor{pp}{JOIN} Country \boldpp{WHERE} surfacearea!=\textbf{\textquotesingle value\textquotesingle}}}  \\
     Top-5 SQL: & \makecell[l]{{\fontfamily{pcr}\selectfont \boldpp{SELECT} \textbf{code} \boldpp{FROM} \textbf{CountryLanguage} \textcolor{pp}{JOIN} Country \boldpp{WHERE} countrycode!=\textbf{\textquotesingle value\textquotesingle}}} 
     \\[-2.0ex] \\ \Xhline{3\arrayrulewidth}
  \end{tabular}
  \end{adjustbox}
  \vspace{-1mm}
  \captionsetup{font=small}
  \caption{An NL-SQL pair and the corresponding translation results of an NLIDB model, with the duplicated parts highlighted in \textbf{bold}.}
  \label{fig:(b)(c)spider query}
 \end{subfigure}
 \captionsetup{font=small}
 \caption{An example from the \textsc{Spider} benchmark}
 \label{fig:spider query}
\end{figure}

 One plausible reason we believe is that using standard auto-regressive decoding to generate single SQL queries may result in sub-optimal outputs in two main aspects: (1) \textbf{Lack of output diversity}. Auto-regressive decoding, commonly used with beam search or sampling methods such as top-K sampling \cite{topk18}, often struggles with generating a diverse set of candidate sequences and tends to exhibit repetitiveness in its outputs \cite{GimpelBDS13, LiGBGD16, LiJ16}. Consider the example in Fig.~\ref{fig:spider query} that shows an NL query with the corresponding translation results of the translation model \textsc{Lgesql} \cite{LGESQL20}. While \textsc{Lgesql} model using beam search maintains a list of top-K best candidates, these outputs are near-duplicates with minor variations, resulting in the final incorrect translation\footnote{Given that a country may have multiple languages spoken, the top-1 translated SQL is considered as incorrect translation, as the country ``\textit{Aruba}'' may be mistakenly selected for the given scenario.}. (2) \textbf{Lack of global context awareness}. Due to the incremental nature of generating output tokens one by one based on the previously generated tokens, auto-regressive decoding may lead to local optima outputs as it considers only partial context \cite{SummaReranker22, PairReranker22, Joint23}, thereby causing a failure to find the correct translation as well.

 To improve the existing end-to-end translation paradigm, a multi-task generation framework \cite{MIGA22} targeting the conversational translation scenario is introduced to improve existing translation models. Although the approach achieves state-of-the-art performance on conversational benchmarks, the framework still relies on the standard auto-regressive decoding procedure to obtain the final results, which may not be optimal. Another recent work \cite{GAR23, GenSql23} proposes a generative approach for the NL2SQL problem, but it requires a hypothesis based on a set of representative sample queries.

 In this paper, we present \textsc{Metasql}, a novel approach aimed at enhancing the auto-regressive decoding process in NL2SQL translation. Drawing inspiration from \textit{controllable text generation} techniques \cite{Controllable23, Controlled23} in NLP, \textsc{Metasql} incorporates control signals \cite{Controllable22}, either explicitly or implicitly, into the standard auto-regressive decoding process, thereby facilitating more targeted SQL generation. To tackle the problem of insufficient output diversity, \textsc{Metasql} introduces \textit{query metadata as an explicit control signal} to manipulate the behavior of translation models for better SQL query candidate generation. Additionally, to overcome the lack of global context, \textsc{Metasql} reframes the NL2SQL problem as a post-processing \textit{ranking procedure (as an implicit control signal)}, leveraging the entire global context rather than partial information involved in sequence generation. Here, the query metadata we mean represents a set of semantic units of a SQL query that serve as generation constraints for constructing the complete SQL query under a specific database. (More details can be found in Section \ref{metadata}.)
 
 Concretely speaking, \textsc{Metasql} introduces a \textbf{unified generate-then-rank framework} that is compatible with any existing Seq2seq-based and LLM-based NL2SQL models to enhance their translation accuracy. Motivated by the recent achievements of task decomposition \cite{Gupta18, Talmor18, decomposition19, Multi-hop19, Break20} and diverse decoding \cite{GimpelBDS13, LiGBGD16, LiJ16} techniques, \textsc{Metasql} incorporates query metadata to upgrade the end-to-end sequence generation paradigm as follows. \circled{1} To understand the given NL query, \textsc{Metasql} first maps the meaning of the NL query into a small set of related query metadata; \circled{2} Next, by manipulating the Seq2seq translation model behavior, \textsc{Metasql} generates a diverse collection of candidate SQL queries by conditioning on different compositions of the retrieved query metadata; \circled{3} Finally, \textsc{Metasql} implements a two-stage ranking pipeline to find the best-matching SQL query as the translation result. \textit{Here, since the ranking pipeline has global information about what the target SQL query to be generated might be, we posit that it has the potential to do a better translation than the traditional left-to-right fashion generation.}
 
 

 To assess the efficiency of \textsc{Metasql}, we conduct our experiments on two public NLIDB benchmarks, namely \textsc{Spider} and \textsc{ScienceBenchmark} \cite{ScienceBenchmark23}, by applying \textsc{Metasql} to four Seq2seq models, \textsc{Bridge} \cite{BRIDGE20}, \textsc{Gap} \cite{GAP21}, \textsc{Lgesql}, and \textsc{Resdsql} \cite{resdsql23}, along with two LLMs, \textsc{Gpt-3.5-turbo} (the model used behind \textsc{ChatGPT}\footnote{\url{https://chat.openai.com}}) and \textsc{Gpt-4}. Experimental results reveal that \textsc{Metasql} consistently enhances the performance of all models across two benchmarks, with \textsc{Lgesql} achieving a translation accuracy of 77.4\% on the validation set and 72.3\% on the test set of \textsc{Spider}, and \textsc{Gpt-4} attaining translation accuracies of 68.6\%, 42.0\% and 17.6\% on the three scientific databases of \textsc{ScienceBenchmark}, respectively.


 To summarize, our contributions are three-fold:
\begin{itemize}[leftmargin=*]
  \item We propose \textsc{Metasql}, a unified framework for the NL2SQL problem, designed to enhance the performance of existing Seq2seq-based and LLM-based translation models.

 \item \textsc{Metasql} formulates the NL2SQL task as a diverse generation and ranking problem by incorporating query metadata to control the generation of better SQL query candidates and utilizing learning-to-rank algorithms to achieve the ranking procedure, thereby enhancing SQL query translation.
 
 \item We perform a series of experiments to evaluate \textsc{Metasql} on two public NLIDB benchmarks with four state-of-the-art Seq2seq-based translation models and two LLMs. The experiments demonstrate the effectiveness of \textsc{Metasql}.

\end{itemize}

 
The remainder of this paper is organized as follows. We first present the overview of \textsc{Metasql} in Section \ref{section 2}; We then go into the details of the methodologies in Section \ref{section 3}. We report the experimental results in Section \ref{section 4}. Finally, we discuss related works in Section \ref{section 5} and conclude in Section \ref{section 6}.

\section{\textsc{Metasql}}\label{section 2}

We first give some essential preliminaries of \textsc{Metasql} and then describe the overall of our approach.

\begin{figure*}[ht]
    \centering
    \includegraphics[width=\linewidth]{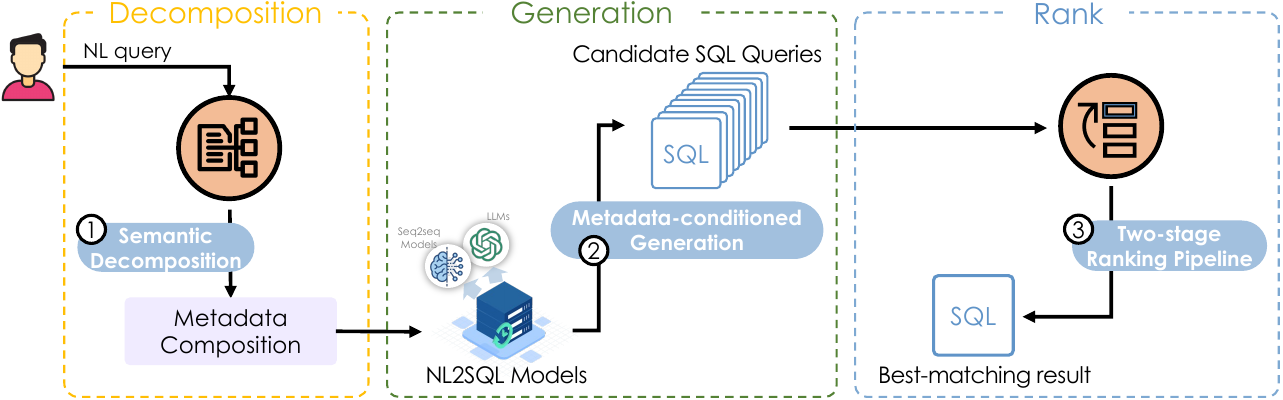}
    \captionsetup{font=small}
    \caption{Overview of \textsc{Metasql}}
    \label{fig:overview of MetaSQL}
\end{figure*}

\subsection{Preliminaries}
\noindent\textbf{Auto-regressive Decoding} refers to a decoding strategy where a model generates output sequences one element at a time, and the generation of each element depends on the previously generated ones. Decoding in auto-regressive models involves learning a scoring model $p(y \vert x)$ that decomposes based on left-to-right factorization,

\begin{equation*}
\textrm{log}(y \vert x) = \sum^{m-1}_{j=0}\textrm{log}p(y_{j+1} \vert y_{\leq j}, x)
\end{equation*}

\noindent where the objective is to find a high-scoring output sequence $y=(y_{1}, \cdots, y_{m}$) given an input sequence $x=(x_{1}, \cdots, x_{n}$). It's worth noting that standard uni-directional decoding algorithms, like greedy and beam search, are ineffective in producing high-scoring output sequences. This inefficiency arises because errors in the decoding history can have adverse effects on future outcomes. These algorithms rely on making local decisions to extend an incomplete sequence by selecting the token with the maximum likelihood at each time step, hoping to achieve a globally optimal complete sequence \cite{BahdanauCB14}.




\noindent\textbf{NL2SQL Models} convert human-readable NL queries into executable SQL queries, which mainly fall into two categories:

\noindent\textit{Seq2seq-based NL2SQL Models.} A Seq2seq-based NL2SQL translation model commonly follows the Seq2seq learning framework \cite{seq2seq14} to translate NL queries to their SQL counterparts. Given an input NL query $\mathcal{X} = \{x_{1},x_{2},\cdots,x_{n}\}$ and a database schema $\mathcal{S}=\left\langle\mathcal{C},\mathcal{T},\mathcal{F}\right\rangle$ that consists of columns $\mathcal{C}=\{c_{1},c_{2},\cdots,c_{\mid\mathcal{C}\mid}\}$, tables $\mathcal{T}=\{t_{1},t_{2},\cdots,t_{\mid\mathcal{T}\mid}\}$, and a set of foreign-primary key pairs $\mathcal{F}=\{(c_{f_{1}},$
$c_{p_{1}}),(c_{f_{2}},c_{p_{2}}),\cdots,(c_{f_{\mid\mathcal{C}\mid}},c_{p_{\mid\mathcal{C}\mid}})\}$, the model uses an encoder to compute a contextual representation $\textbf{\em c}$ by jointly embedding the NL query $\mathcal{X}$ with schema $\mathcal{S}$. Afterward, an auto-regressive decoder is used to compute a distribution $P(\mathcal{Y}\mid\textbf{\em c})$ over the SQL programs $\mathcal{Y}=(y_{1},\cdots,y_{m})$. Depending on different model designs, the learning target $\mathcal{Y}$ of the decoder can be raw SQL tokens \cite{EditSQL19,BRIDGE20}, intermediate representations of SQL language \cite{IRNet19, NatSQL21}, or SQL abstract syntax trees \cite{RATSQL20, LGESQL20, GAP21}.

\noindent\textit{LLMs as NL2SQL Models.} In light of the recent advancements in LLMs, current research \cite{nitarshan22, aiwei23} endeavors to employ LLMs as NL2SQL models without fine-tuning. By providing an NL query $\mathcal{X}$ and a prompt $\mathcal{P}$ as input, an LLM can be utilized to auto-regressively generate the corresponding SQL query $\mathcal{Y}$, akin to the decoding process of Seq2seq-based translation models. Depending on the prompting technique utilized, such as zero-shot, few-shot prompting, or in-context learning, the prompt $\mathcal{P}$ can include text instructions \cite{aiwei23}, translation demonstrations \cite{SQL-PaLM23} or reasoning chains \cite{DINSQL23, Interleaving23}.

\subsection{Overview}
A high-level view of \textsc{Metasql} can be seen in Fig.~\ref{fig:overview of MetaSQL}. The main process is as follows:

\begin{enumerate}
 \item NL query semantic parsing is reformulated as a classification problem, where the NL semantics are mapped to a set of related query metadata using a multi-label classifier.
 \item (Optional) An underlying translation model is supervised-trained on augmented NL-SQL data with additional metadata added to the NL part.
 \item Conditioned on different compositions of the related query metadata for the given NL query, a set of diverse candidate SQL queries is then generated by using the translation model.
 \item A two-stage ranking pipeline is applied to get a global-optimal SQL query as the translation result based on the semantic similarity with the given NL query.
\end{enumerate}

\noindent Among these, the metadata-conditioned generation followed by ranking is unique to our setup, and we found that this process is the key to improving translation accuracy. We describe each above step below.

\noindent\textbf{Semantic Decomposition.} This step in Fig.~\ref{fig:overview of MetaSQL}-\circled{1} is to decompose the meaning of the given NL query and map it to a set of query metadata. This is accomplished by treating the semantic decomposition task as a classification problem and using a multi-label classifier to select all relevant query metadata for the given NL query. Here, the query metadata is represented as a set of categorical labels that capture the context expressed by the NL query in relation to the underlying database. For example, suppose the NL query in Fig.~\ref{fig:(b)(c)spider query} is given. In that case, \textsc{Metasql} should select ``\textit{project}'' and ``\textit{except}'' query operator labels, along with a hardness value ``\textit{400}'' indicating the SQL query's anticipated difficulty level as corresponding metadata. (A detailed definition can be found in Section \ref{section 3.1}.)

\noindent\textbf{Metadata-conditioned Generation.}\label{section: annotation} This step in Fig.~\ref{fig:overview of MetaSQL}-\circled{2} is to employ the base translation model to produce a list of candidate SQL queries for the given NL query. \textsc{Metasql} achieves this by manipulating the behavior of the translation model to generate a collection of SQL queries by conditioning on different compositions of the retrieved metadata from the last step. Continuing the running example in Fig.~\ref{fig:spider query}, the following SQL query is one of the candidate queries that may be generated by the translation model conditioned on the ``\textit{where}'' label with a rating value ``200'':

\vspace{2mm}
\noindent {\fontfamily{pcr}\fontsize{9}{10}\selectfont \textcolor{pp}{\textbf{SELECT}} countrycode \textcolor{pp}{\textbf{FROM}} CountryLanguage}  \\
{\fontfamily{pcr}\fontsize{9}{10}\selectfont \textcolor{pp}{\textbf{WHERE}} language!=\textcolor{rd}{\textquotesingle English\textquotesingle}}
\vspace{2mm}

\noindent Note that in order to use the query metadata for translation assistance, the conventional supervised learning process of Seq2seq translation models requires enhancement through the inclusion of the metadata into the model input. However, since LLMs can serve as NL2SQL models effectively without requiring fine-tuning, the training procedure is not required.

\noindent\textbf{Two-stage Ranking Pipeline.} This step in Fig.~\ref{fig:overview of MetaSQL}-\circled{3} is to utilize a ranking procedure to determine which candidate SQL query is the correct translation to a given NL query. Inspired by the recent success of the multiple-stage ranking paradigm in information retrieval \cite{multi-bert19, Gao21}, \textsc{Metasql} utilizes a two-stage ranking pipeline. In this pipeline, the initial ranking stage produces a set of more relevant candidates for the second-stage ranking model to identify the top-ranked query. Here, the ranking models learn to rank the semantic similarity across two modalities (i.e., NL and SQL). For example, for the given NL query in Fig.~\ref{fig:(b)(c)spider query}, the ranking pipeline recognizes the below ground-truth SQL query as the most similar query for the given NL query and hence the translation result.

\vspace{2mm}
\noindent {\fontfamily{pcr}\fontsize{9}{10}\selectfont \textcolor{pp}{\textbf{SELECT}} countrycode \textcolor{pp}{\textbf{FROM}} CountryLanguage} \\
{\fontfamily{pcr}\fontsize{9}{10}\selectfont \textcolor{pp}{\textbf{EXCEPT}} \textcolor{pp}{\textbf{SELECT}} countrycode \textcolor{pp}{\textbf{FROM}}} \\
{\fontfamily{pcr}\fontsize{9}{10}\selectfont  CountryLanguage \textcolor{pp}{\textbf{WHERE}} language=\textcolor{rd}{\textquotesingle English\textquotesingle}}
\vspace{2mm}

The training data of the ranking models are composed of a set of triples $\{(q_{i}, s_{i}, y_{i}) \vert q_{i}\in Q, s_{i}\in S, 0 \leq y_{i}\leq 10\}_{i=1}^{N}$, where $q_{i}$ represents an NL query, $s_{i}$ denotes a SQL query, and $y_{i}$ represents the semantic similarity score between $s_{i}$ and $q_{i}$, such that the more similar $s_{i}$ and $q_{i}$ are, the closer the score $y_{i}$ is to $10$. In this paper, the semantic similarity score $y_{i}$ is set to $10$ if $s_{i}$ is exactly the ``gold'' SQL query of $q_{i}$. Otherwise, $y_{i}$ is calculated by comparing each clause of $s_{i}$ with the given ``gold'' SQL query for $q_{i}$.

\section{Methodologies}\label{section 3}
\label{section:methodologies}

 In this section, we first elaborate on the query metadata design, then describe in detail the metadata selection, metadata-conditioned generation, and two-stage ranking of \textsc{Metasql}.

\subsection{Query metadata}\label{section 3.1}

\subsubsection{Metadata Design}\label{metadata} We design the query metadata to be expressive enough to represent the high-level semantics that the query (NL and SQL counterpart) may express. In \textsc{Metasql}, we introduce the following three types of metadata - \textit{operator tag}, \textit{hardness value} and \textit{correctness indicator}. 


\begin{itemize}[leftmargin=*]
 \item \textbf{Operator Tag.} Each operator tag corresponds to a single logical operator, where each operator either selects a set of entities, retrieves information about their attributes, or aggregates information over entities. Note that since the operators are primarily inspired by SQL, this kind of metadata indicates which SQL components should be used for translating the given NL query. 
 
 \noindent For example, as for the NL query in Fig.~\ref{fig:spider query}, the query should correspond to ``\textit{project}'' and ``\textit{except}'' operator tags.
 \item \textbf{Hardness Value.} Hardness value serves as a metric to quantify the potential complexity of a query. This definition draws from the \textit{SQL hardness criteria} outlined in \cite{Yu18}, where query complexity is assessed based on the number and type of SQL components present in a query. Taking an additional step, we utilize the criteria to assign a difficulty score to each SQL component (with a base value of $50$), reflecting its syntactic complexity. The hardness value for a given query is then determined by summing the difficulty scores of its individual SQL components.
 
    
 \noindent In the example of Fig.~\ref{fig:spider query}, the hardness value of the query is set to 400, according to the initial rating of 100 combined with the ``hard'' {\fontfamily{qcr}\selectfont \textcolor{pp} {\textbf{EXCEPT}}} clause rating of 300. 
    
 \item \textbf{Correctness Indicator.} We use a correctness indicator to distinguish correct queries from incorrect ones. Note that this kind of metadata is always true in both the semantic decomposition and metadata-conditioned generation steps at the inference time of \textsc{Metasql}, but be changeable with either ``\textit{correct}'' or ``\textit{incorrect}'' at training time for model learning. More details about the usage of this kind of metadata can be found in Section \ref{generation}.
\end{itemize}


\subsubsection{Semantic Decomposition with Metadata}\label{multi-label} To retrieve metadata from a given NL query, we frame the NL-to-metadata mapping as a classification problem. Here, the metadata is treated as a collection of categorical values, with each individual metadata value representing a distinct class. We utilize a multi-label classification model to implement this mapping.

Technically, the architecture of the multi-label classifier can be derived from any underlying NL2SQL translation model, in the sense that they share the same encoder, but the decoder is replaced with a classification layer to output scalar values. In this manner, the multi-label classification model benefits from the encoding capacities maintained in the translation model. More specifically, the classifier reuses the encoder of a translation model to encode the NL query and the corresponding database schema jointly. Then, the contextual representation is then fed into a classification layer to calculate the possibility mass over different categorical values of query metadata. We set a classification threshold of $p$ to select the possible metadata labels with a higher probability mass over all outputs at the inference stage.

By deconstructing the semantics of the given NL query into its corresponding set of query metadata, \textsc{Metasql} can capture more fine-grained semantics, allowing the discovery of diverse semantic-equivalent SQL queries based on various combinations of plausible query metadata.

\subsection{Metadata-conditioned Generation}\label{generation}
An essential question for \textsc{Metasql} is how the metadata information can be enforced in the sequence generation process of the traditional Seq2seq-based translation models. \textsc{Metasql} is primarily inspired by the prompting methods \cite{Austin21, Cobbe21, AlphaCode22} and takes it further by incorporating query metadata as additional language prompts to enhance the sequence generation.

In the rest of this section, we first elaborate on how \textsc{Metasql} trains Seq2seq-based NL2SQL models with query metadata as additional language prompts and then explain the metadata-conditioned generation process of these Seq2seq NL2SQL models (and LLMs) during the inference stage.

\subsubsection{Training with Metadata} In model training, we add query metadata as prefix language prompts to the NL queries and follow the traditional seq2seq paradigm. The metadata provides an additional learning signal, alleviating the burden of parsing complex queries for the model.

\noindent\textbf{Training Data.} \textsc{Metasql} collects the training data for the underlying NL2SQL model using the principle of weak supervision. Specifically, the training data of the translation model is enforced as a set of triples $\{(q_{i}, s_{i}, M_{i})\}_{i=1}^{N}$, where $q_{i}$ is an NL query, $s_{i}$ is the corresponding SQL query and $M_{i}$ is the query metadata associated with a given NL query. We collect the query metadata $M_{i}$ as follows: Firstly, for the operation tag-type metadata, we directly examine the corresponding SQL query $s_{i}$ and get the relevant operation tags. Secondly, for the hardness value-type metadata, we use the definition of hardness used in the \textsc{Spider} \cite{Yu18} benchmark and assign scores to each syntactical structure in an SQL query to calculate the value. Lastly, we determine the correctness indicator-type metadata based on the data types. Namely, if the data originates from the public NLIDB benchmarks, we consider it a positive sample; otherwise, we label it a negative sample.

\noindent\textbf{Negative Samples.} To allow Seq2seq translation models to better differentiate between correct and incorrect target sequences (i.e., SQL queries), we gather the erroneous translations from existing translation models on the training set of the \textsc{Spider} benchmark and use these translations as negative samples to augment the training data. Hence, we assign the ``\textit{incorrect}'' correctness indicator as part of the query metadata for these negative samples. By doing this, translation models may intentionally circumvent the wrong parsing path by using this type of metadata during the learning process.

\begin{figure}[ht]
    \centering
    \includegraphics[width=\linewidth]{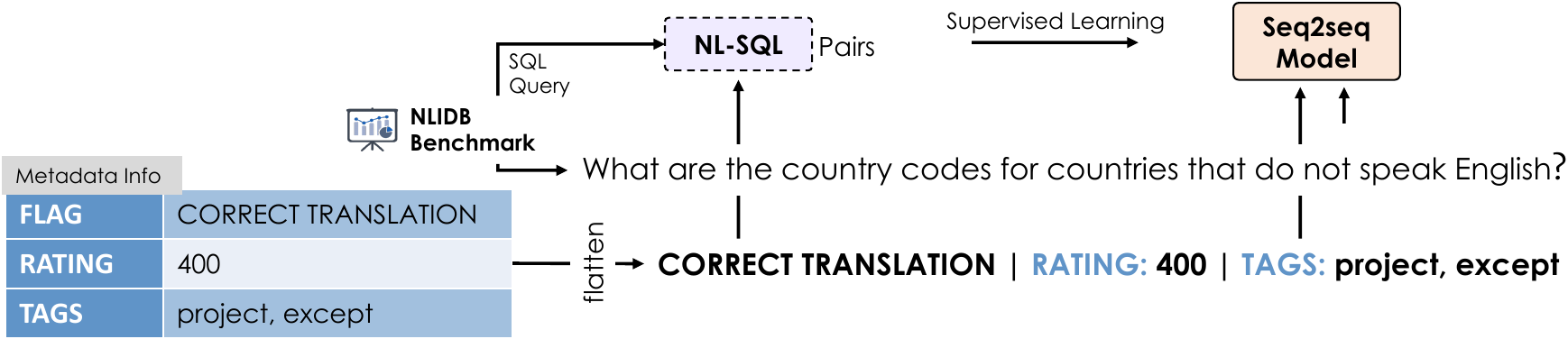}
    \captionsetup{font=small}
    \caption{Training procedure of Seq2seq-based models in \textsc{Metasql}.}
    \label{fig:model input}
\end{figure}

\noindent\textbf{Model Input.} As illustrated in Fig.~\ref{fig:model input}, the translation model input comprises both the NL query and its associated query metadata. To add query metadata as a prefix, we flatten the metadata into a sequence and then concatenate it with the given NL query. All metadata, including the hardness value, is treated as a string. For example, a flattened query metadata $M$ for the NL query in Fig.~\ref{fig:(b)(c)spider query} is represented below:

$M = correct\;|\;rating:400\;|\;tags:project,except$

\noindent Here $|$ is a special token to separate different metadata. This allows us to prefix the flattened metadata $M$ with the NL query $q$ before feeding it into the encoder of the translation model.

\subsubsection{Conditioned Generation.} As the query metadata is unknown during inference time, \textsc{Metasql} utilizes the multi-label classification model introduced in the previous Section \ref{multi-label} to obtain the query metadata, thus diversifying plausible translations by conditioning on the metadata.

More precisely, given an NL query, \textsc{Metasql} initially employs the multi-label classifier to obtain an initial set of metadata labels. To ensure the controlled sampling of semantically relevant query metadata conditions, \textsc{Metasql} selectively composes these labels by considering combinations observed in the training data, assuming that the training and test data share the same distribution. Using the resulting sampled metadata conditions, \textsc{Metasql} manipulates the behavior of the trained model and generates a set of candidate SQL queries by conditioning each query metadata condition. This approach is similar to the prompt-based methods used in LLMs \cite{GPT3, GPT4} to generate textual responses to different given prompts defined by specific downstream tasks.

\begin{figure}[ht]
    \centering
    \includegraphics[width=\linewidth]{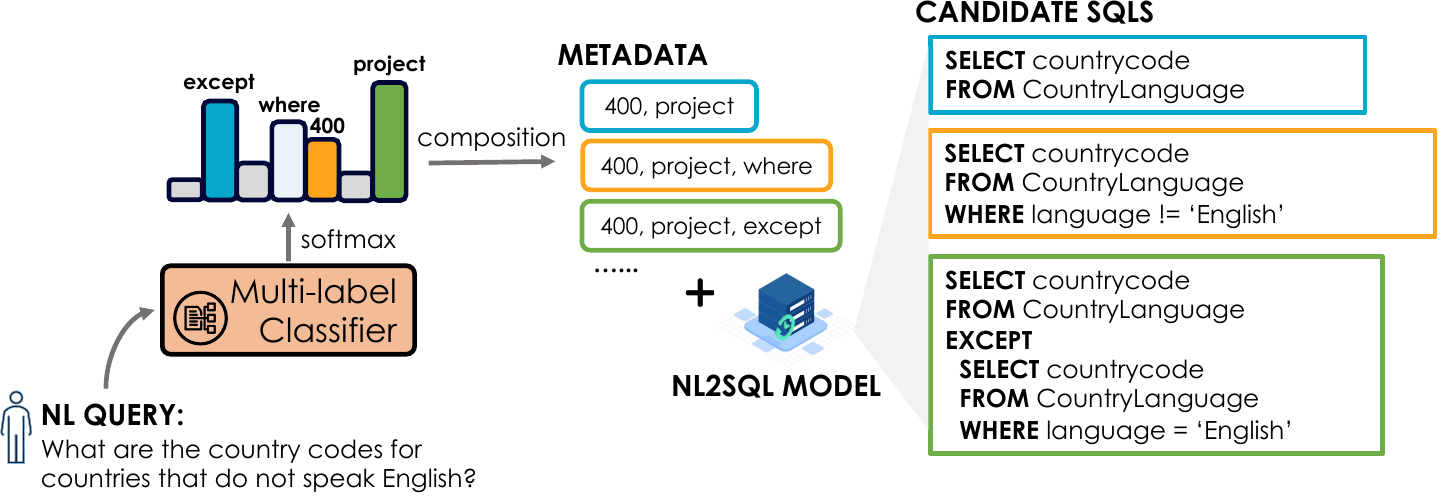}
    \captionsetup{font=small}
    \caption{Metadata-conditioned generation in \textsc{Metasql}.}
    \label{fig:sampling}
\end{figure}

For the NL query example in Fig.~\ref{fig:spider query}, we illustrate the generation results of \textsc{Metasql} during the inference time, as depicted in Fig.~\ref{fig:sampling}. Using metadata labels ``\textit{400}'', ``\textit{project}'', ``\textit{where}'', and ``\textit{except}'' obtained from the multi-label classification model, we derive three distinct compositions of query metadata and generate three distinct SQL queries based on each query metadata, respectively\footnote{Note that each operator tag-type metadata indicates the presence of a specific SQL operation without limiting the number of attributes used in it. Hence, multiple projections-SQL can be generated in the given example.}.



\subsection{Two-stage Ranking Pipeline}\label{reranking section}
 \textsc{Metasql} implements a ranking pipeline with two separate machine learning-based ranking models across two modalities (i.e., NL and SQL). In the first stage, a coarse-grained ranking model narrows the relatively large set to a relatively small collection of potential candidates. Then, a second-stage fine-grained ranking model is applied to the resulting set from the first stage to get the final top-ranked SQL query.

\begin{figure}
  \begin{subfigure}[b]{1.0\linewidth}
   \centering
   \includegraphics[width=.7\linewidth]{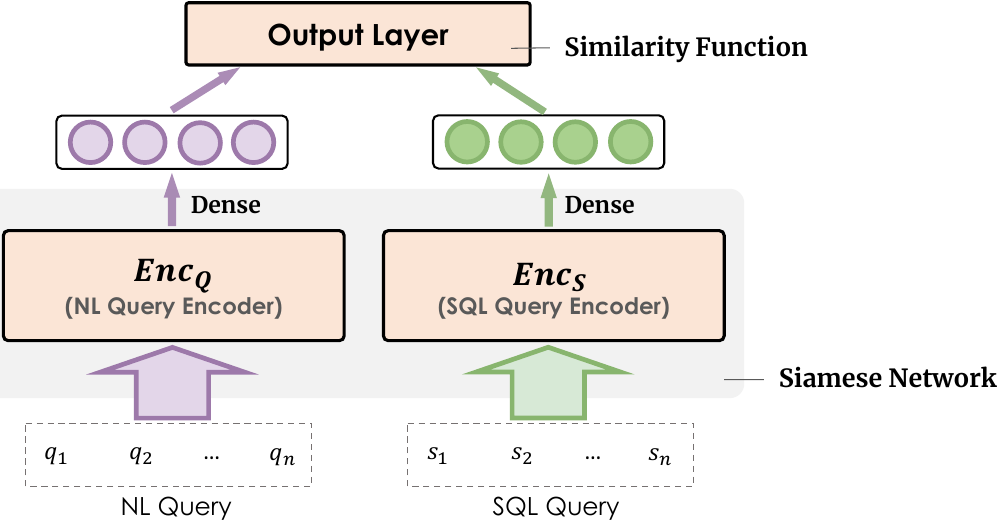}
   \captionsetup{font={small}}
   \caption{The dual-tower architecture used in first-stage ranking model.}
   \label{fig:first-stage model}
  \end{subfigure}
  \begin{subfigure}[b]{1.0\linewidth}
    \includegraphics[width=\linewidth]{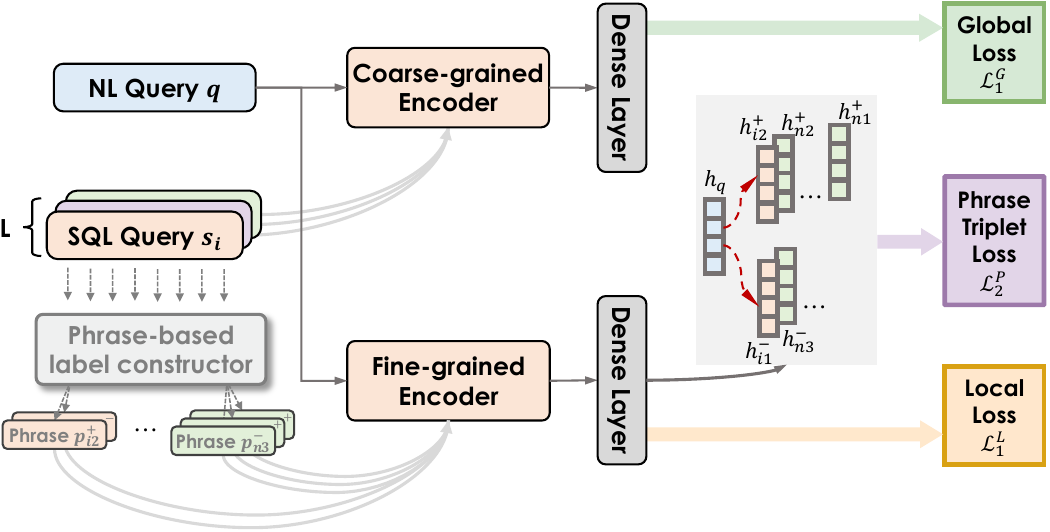}
    \captionsetup{font=small}
    \caption{Second-stage ranking with the multi-grained supervision signals.}
    \label{fig:multi-grained}
  \end{subfigure}
  \captionsetup{font=small}
  \caption{Two-stage ranking models used in \textsc{Metasql}.}
\end{figure}

\subsubsection{First-stage Ranking Model} 
To fast retrieve a relatively small set of candidate SQL queries, we employ the widely used \textit{dual-tower architecture} \cite{LeeCT19, KarpukhinOMLWEC20, XiongXLTLBAO21} in information retrieval to construct the first-stage ranking model. Fig.~\ref{fig:first-stage model} presents the overall architecture of the first-stage ranking model. Specifically, the network architecture includes two \textsc{Bert}-like bidirectional text encoders \cite{BERT19} (i.e., an NL query encoder $\mathrm{Enc}_{Q}$ and a SQL encoder $\mathrm{Enc}_{S}$) and uses the cosine function as the similarity function to measure the semantic similarity between the NL query and the SQL query as follows:

\begin{equation}
    \mathrm{sim}(q, s) = \frac{\mathrm{Enc}_{Q}(q) \cdot \mathrm{Enc}_{S}(s)}{\parallel \mathrm{Enc}_{Q}(q) \parallel \parallel \mathrm{Enc}_{S}(s)\parallel} 
\end{equation}



\noindent\textbf{Training Data.} The training data of the first-stage ranking model is a set of triples $\{(q_{i}, s_{i}, v_{i} )\}_{i=1}^{N}$, where $q_{i}$ is an NL query, $s_{i}$ is a SQL query and $v_{i}$ is the semantic similarity score between $q_{i}$ and $s_{i}$. The score $v_{i}$ is determined as follows: If $s_{i}$ corresponds to the ``gold'' SQL query of $q_{i}$, then $s_{i}$ is set to $1$. Such a triple is called a positive sample, which is obtainable from public benchmarks. Otherwise, $v_{i}$ is calculated by comparing each clause in the SQL query $s_{i}$ with the corresponding ``gold'' SQL query for $q_{i}$. If a clause differs, a penalty is applied to the $v_{i}$ value. The calculation continues until all the clauses are compared or $v_{i}$ drops to $0$. Such a triple is called a negative sample, which can be collected using the trained Seq2seq-based translation model in Section \ref{generation} by conditioning on different metadata.

 

\subsubsection{Second-stage Ranking Model} 
\label{re-ranking model}
The objective of the re-ranking model is to accurately find the top-ranked SQL query based on the semantic similarities with the given NL query from the resulting set from the first-stage ranking model. Nevertheless, we observe that most current ranking architectures, like the one used in the first stage, primarily rely on \textit{sentence-level supervision} to distinguish matched and mismatched candidates, which is limited for a precise ranking purpose. Table \ref{tab:fine grain example} presents an example of the ranking results produced by the first-stage ranking model. Mismatched sentences are usually partially irrelevant with phrases of inconsistent semantics (the missed {\fontfamily{qcr}\selectfont \textcolor{pp}{\textbf{WHERE}} pets.pettype=\textcolor{rd}{\textquotesingle cat\textquotesingle}}, the mismatched {\fontfamily{qcr}\selectfont \textcolor{pp}{\textbf{JOIN}} pets}, etc.). This example shows that the semantic mismatch usually happens in finer grain, i.e., phrase level.

\setul{}{2pt}
\definecolor{grey}{rgb}{0.70, 0.70, 0.70}
\begin{table}[ht]
  \centering
  \captionsetup{font=small}
  \caption{An example of an NL query, a group of mismatched SQL queries, and the corresponding matched SQL query. Query segments with underlines stand for mismatching at phrase level.}
  \vspace{-2mm}
  \begin{adjustbox}{width=\linewidth}
  \begin{tabular}{|c|l|c|} \Xhline{3\arrayrulewidth}
     \textbf{NL Query} & \makecell[l]{\emph{Find the last name of the student who has a cat that is age 3.}} & \textbf{Similarity Score}  \\ \hline

     & \makecell[l]{{\fontfamily{pcr}\selectfont \textcolor{pp}{\textbf{SELECT}} student.lname} \\ {\fontfamily{pcr}\selectfont \textcolor{pp}{\textbf{FROM}} student \textcolor{pp}{\textbf{JOIN}} has\_pet \textcolor{pp}{\textbf{JOIN}} pets} \\ {\fontfamily{pcr}\selectfont \textcolor{pp}{\textbf{WHERE}} pets.pet\_age=\textcolor{rd}{3} \textcolor{grey}{\ul{\textbf{AND} pets.pettype=\textquotesingle cat\textquotesingle}}} \\ [2.0ex]} & 0.76 \\
     
     \textbf{\makecell[c]{Mismatched \\ SQL Queries}} & \makecell[l]{{\fontfamily{pcr}\selectfont \textcolor{pp}{\textbf{SELECT}} student.lname} \\ {\fontfamily{pcr}\selectfont \textcolor{pp}{\textbf{FROM}} student \textcolor{grey}{\ul{\textbf{JOIN} has\_pet}} \textbf{JOIN} pets} \\ {\fontfamily{pcr}\selectfont \textcolor{pp}{\textbf{WHERE}} pets.pettype=\textcolor{rd}{\textquotesingle cat\textquotesingle} \textcolor{pp}{\textbf{AND}} pets.pet\_age=\textcolor{rd}{3}} \\ [2.0ex]} & 0.82 \\
     
     & \makecell[l]{{\fontfamily{pcr}\selectfont \textcolor{pp}{\textbf{SELECT}} student.lname, \ul{pets.pettype}} \\ {\fontfamily{pcr}\selectfont \textcolor{pp}{\textbf{FROM}} student \textcolor{pp}{\textbf{JOIN}} has\_pet \textcolor{pp}{\textbf{JOIN}} pets} \\ {\fontfamily{pcr}\selectfont \textcolor{pp}{\textbf{WHERE}} pets.pet\_age=\textcolor{rd}{3} \textcolor{pp}{\textbf{AND}} pets.pettype=\textcolor{rd}{\textquotesingle cat\textquotesingle}} \\ [1.5ex]} & 0.73 \\ \hline
     
     \textbf{\makecell[c]{Matched \\ SQL Query}} & \makecell[l]{{\fontfamily{pcr}\selectfont \textcolor{pp}{\textbf{SELECT}} student.lname} \\ {\fontfamily{pcr}\selectfont \textcolor{pp}{\textbf{FROM}} student \textcolor{pp}{\textbf{JOIN}} has\_pet \textcolor{pp}{\textbf{JOIN}} pets} \\ {\fontfamily{pcr}\selectfont \textcolor{pp}{\textbf{WHERE}} pets.pettype=\textcolor{rd}{\textquotesingle cat\textquotesingle} \textcolor{pp}{\textbf{AND}} pets.pet\_age=\textcolor{rd}{3}} \\ [1.5ex]} & 0.72 \\ \Xhline{3\arrayrulewidth}
  \end{tabular}
  \end{adjustbox}
  \label{tab:fine grain example}
 \end{table}

Motivated by this finding and the recent advancements in image-text retrieval \cite{LiuMZX0Z20, multigrained22}, we explore providing multi-grained supervision signals (i.e., incorporating both \textit{\textbf{sentence-level}} and \textit{\textbf{phrase-level}} supervision) in the second-stage ranking model for better identification of mismatched components in the SQL queries. Fig.~\ref{fig:multi-grained} presents the architecture of our proposed second-stage ranking model, which includes two encoders (i.e., the upper coarse-grained encoder and the lower fine-grained encoder) for multi-grained semantics capture. Note that we employ the \textit{listwise approach} \cite{listwise07} to construct the second-stage ranking model. That is, the training setting of the second-stage ranking model consists of a finite dataset consisting of $n$ triplets $D = \{q_{i}, S_{i}, Y_{i}\}_{i=1}^{N}$, where $S_{i}=\{s_{i, 1}, s_{i, 2}, \cdots, s_{i, L}\}$ is the list of SQL queries, and $Y_{i}=\{y_{i, 1}, y_{i, 2}, \cdots, y_{i, L}\}$ are the corresponding relevance similarity scores of $S_{i}$.


\noindent\textbf{Multi-Grained Feature Construction.} To capture the semantics of different granularities for a given SQL query, we introduce additional phrase-level semantics from its original form. Drawing inspiration from query translation studies \cite{SQL2NL10, Iyer16, Xu18}, we adopt a straightforward \textit{rule-based approach} to systematically generate an NL description for a specific SQL unit. This involves linking each type of SQL unit with a pre-determined template, then populated with element-based labels extracted from the SQL unit to form the NL description. The different types of SQL units used in \textsc{Metasql} are listed in Table \ref{tab:component types}, and more details can be seen in \cite{SQL2NL10}.

\begin{table}[ht]
\centering
\captionsetup{font=small}
\caption{Query unit types and examples}
    \vspace{-2mm}
    \begin{adjustbox}{width=\linewidth}
    \begin{tabular}{|c | c | c|}
        \hline
        \textbf{Type} & \textbf{Unit Example} & \textbf{NL Description} \\ \hline
        \textsc{Projection} & \multicolumn{1}{l|}{\fontfamily{qcr}\fontsize{8}{9}\selectfont \textcolor{pp}{\textbf{SELECT}} employee.name} & \multicolumn{1}{l|}{Find the employee name.}\\  \hline
        \multirow{2}{*}{\textsc{Join}}& \multicolumn{1}{l|}{{\fontfamily{qcr}\fontsize{8}{9}\selectfont \textcolor{pp}{\textbf{FROM}} employee}} & \multicolumn{1}{l|}{Employee} \\ \cline{2-3}
         & \multicolumn{1}{l|}{\makecell[l]{\fontfamily{qcr}\fontsize{8}{9}\selectfont \textcolor{pp}{\textbf{FROM}} employee \textcolor{pp}{\textbf{JOIN}} evaluation \\ \fontfamily{qcr}\fontsize{8}{9}\selectfont  \textcolor{pp}{\textbf{ON}} id=employee\_id}} & \multicolumn{1}{l|}{The employee with evaluation.} \\ \hline
         \multirow{2}{*}{\textsc{Predicate}} & \multicolumn{1}{l|}{{\fontfamily{qcr}\fontsize{8}{9}\selectfont \textcolor{pp}{\textbf{WHERE}} employee.name=\textcolor{rd}{\textquotesingle John\textquotesingle}}} & \multicolumn{1}{l|}{The employee named John.} \\ \cline{2-3}
         & \multicolumn{1}{l|}{\makecell[l]{\fontfamily{qcr}\fontsize{8}{9}\selectfont \textcolor{pp}{\textbf{INTERSECT}} \textcolor{pp}{\textbf{SELECT}} id \textcolor{pp}{\textbf{FROM}}  \\ \fontfamily{qcr}\fontsize{8}{9}\selectfont employee \textcolor{pp}{\textbf{WHERE}} name=\textcolor{rd}{\textquotesingle John\textquotesingle}}} & \multicolumn{1}{l|}{\makecell[l]{(Find the ID of) the employee \\ named John}} \\ \hline
         \textsc{Group} & \multicolumn{1}{l|}{\fontfamily{qcr}\fontsize{8}{9}\selectfont \textcolor{pp}{\textbf{SELECT}} employee.name} & \multicolumn{1}{l|}{Find the employee name.}\\  \hline
         \multirow{2}{*}{\textsc{Sort}} & \multicolumn{1}{l|}{\makecell[l]{\fontfamily{qcr}\fontsize{8}{9}\selectfont \textcolor{pp}{\textbf{ORDER BY}} evaluation.bonus \\ \fontfamily{qcr}\fontsize{8}{9}\selectfont desc \textcolor{pp}{\textbf{LIMIT}} 1}} & \multicolumn{1}{l|}{The highest one time bonus.} \\ \hline
    \end{tabular}
    \end{adjustbox}
    \label{tab:component types}
\end{table}



\noindent\textbf{Multi-scale Loss Construction.} We compute matching scores for NL-SQL pairs using three distinct loss functions: \textit{global}, \textit{local}, and \textit{phrase loss}. Omitting the triplet index, we denote the similarity score vector as $y \in \mathbb{R}^{L}$ and the model’s score vector obtained via the ranking network as $\hat{y^{G}} \in \mathbb{R}^{L}$.

\begin{itemize}[leftmargin=*]
\item\textit{NL-to-SQL Global Loss.} The sentence-level representations of NL and SQL queries measure a global (coarse-gained) cross-modal matching similarity. The loss is shown below,

\begin{equation}
\mathcal{L}_{0}^{G} = \frac{1}{N} \sum^{N}_{i=1}(\hat{y}^{G}_{i}-y_{i})^2
\end{equation}

\noindent where $\hat{y^{G}}$ is the matching scores produced by the coarse-grained encoder with the following dense layer. To align with the listwise paradigm, we further extend the above loss by using listwise NeuralNDCG loss function\cite{Pobrotyn21}.

\item\textit{NL-to-Phrase Local Loss.} We utilize local loss based on NL-to-phrase relationship modeling to enhance the fine-grained cross-modal matching between NL and SQL counterparts. The loss is formulated in the following equation,

\begin{equation}
\mathcal{L}_{1}^{L} = \frac{1}{N} \sum^{N}_{i=1}(\sum^{K}_{k=1}\hat{y}_{i,k}^{L}-y_{i})^2
\end{equation}

\noindent where $\hat{y}_{i,k}^{L}$ is the matching scores produced by the fine-grained encoder with the following dense layer, $K$ denotes the number of phrases produced for a given SQL query. Moreover, we use the listwise version of the loss function to further expand this loss.

\item\textit{Phrase Triplet Loss.} To maximize the fine-grained similarity within a positive pair and minimize the similarity within a negative pair, we split the phrases of candidate SQL queries for a given NL query into a positive set $h_{s_{i}}^{+}$, and a negative set $h_{s_{i}}^{-}$, respectively. Considering that positive parts are the key to separating the mismatched image text pair, we propose $L_{3}^{P}$ to further push away negative parts against positive ones in the negative sentence. It also can be interpreted as the penalty on mismatched parts to guide the matching model to make decisions more grounded on them. We use the triplet loss $TriL_{\alpha}$ to calculate as follows,

\begin{equation}
\mathcal{L}_{3}^{P} = TriL_{\alpha}(h_{q_{i}}, h_{s_{i}}^{+}, h_{s_{i}}^{-})
\end{equation}
\noindent where $\alpha$ is a scalar to regulate the distance between the cosine score of the NL query, positive and negative samples.
\end{itemize}

\noindent\textbf{Inference.} During inference, we use the score $(Q_{i}, S_{i})$ for each item to rank the list of candidate SQL queries,

\begin{equation}
score(q_{i}, s_{i}) = \hat{y}^{G}_{i} + \sum^{K}_{k=1}\hat{y}_{i,k}^{L}
\end{equation}

\section{Experimental Evaluation}\label{section 4}
In this section, we assess the performance of \textsc{Metasql} by applying it to the most advanced NL2SQL models. 



\subsection{Experimental Setup}

 
\subsubsection{Benchmarks}
We conduct extensive experiments on the challenging NLIDB benchmarks \textsc{Spider} and \textsc{ScienceBenchmark} to evaluate the performance of \textsc{Metasql}.

\begin{itemize}[leftmargin=*]
\item\noindent\textsc{Spider} \cite{Yu18} is a large-scale cross-domain benchmark, which includes $10,181$ NL queries and $5,693$ unique SQL queries on 206 databases with multiple tables covering $138$ different domains. \textsc{Spider} authors further split the data into $4$ types, namely \textit{Easy, Medium, Hard, and Extra Hard}, based on the SQL hardness criteria we mentioned in Section \ref{section 3.1}. That is, queries that contain more SQL keywords such as {\fontfamily{qcr}\selectfont\textcolor{pp}{\textbf{GROUP BY}}}, {\fontfamily{qcr}\selectfont\textcolor{pp}{\textbf{INTERSECT}}}, nested subqueries, and aggregators, are considered to be harder.

In light of the inaccessible \textsc{Spider} test set behind an evaluation server, our experiments primarily focused on the \textsc{Spider} validation set. We apply \textsc{Metasql} to \textsc{Lgesql} and submit to the \textsc{Spider} authors to get the evaluation result on the test set of \textsc{Spider} benchmark (See Table \ref{tab:metasql overall accuracy results}).

\item\noindent\textsc{ScienceBenchmark} \cite{ScienceBenchmark23} serves as a complex benchmark for three real-world, scientific databases, namely \textsc{\textit{OncoMx}}, \textsc{\textit{Cordis}} and \textsc{\textit{Sdss}}. For this benchmark, domain experts crafted 103/100/100 high-quality NL-SQL pairs for each domain, then augmented with synthetic data generated using \textsc{Gpt-3}.

In our experiments, we use the \textsc{\textit{Spider}} \textit{Train (Zero-Shot) setting} (i.e., train models on \textsc{Spider} train set, and directly run the evaluation on the human-curated dev set of the respective three databases) introduced in \cite{ScienceBenchmark23}.
\end{itemize}

\definecolor{indigo}{rgb}{0.43, 0.35, 0.83}
\definecolor{sea}{rgb}{0.27, 0.63, 0.64}
\definecolor{dustyorange}{rgb}{0.96, 0.52, 0.13}
\begin{table*}[ht]
  \centering
  \captionsetup{font=small}
  \caption{Illustrate few-shot prompts with LLMs, exclusively applying metadata (underlined) in combination with \textsc{Metasql}.}
  \vspace{-2mm}
  \begin{tabularx}{\textwidth}{cX} 
    \toprule
    \makecell{\textbf{Instruction}} & 
    \makecell[l]{\textbf{\#\#\#\#} Give you database schema, NL question, \ul{and metadata information of the target SQL,} generate an  SQL query.} \\
    \midrule
    \makecell{\textbf{Demonstrations}} & 
    \makecell[l]{\textbf{\#\#\#\#} Learn from the generating examples: \\
    Schema: Table Player with columns \textquotesingle pID\textquotesingle, \textquotesingle pName\textquotesingle, \textquotesingle yCard\textquotesingle, \textquotesingle HS\textquotesingle; Table Tryout with columns \textquotesingle pID\textquotesingle, \textquotesingle cName\textquotesingle, \textquotesingle pPos\textquotesingle, \textquotesingle decision\textquotesingle; \\
     Question: For each position, what is the maximum number of  hours for students who spent more than 1000 hours training?;\\
     \ul{The target SQL only uses the following SQL keywords: JOIN, WHERE, GROUP; The difficulty rating of the target SQL is 350;
     }\\ 
     \textbf{\#\#\#\#} The target SQL is: \\{\fontfamily{qcr}\selectfont \textcolor{pp}{\textbf{SELECT}} \textcolor{blue}{max}(T.HS),T2.pPos \textcolor{pp}{\textbf{FROM}} player \textcolor{pp}{\textbf{AS}} T \textcolor{pp}{\textbf{JOIN}} tryout \textcolor{pp}{\textbf{AS}} T2 \textcolor{pp}{\textbf{WHERE}} T.HS\textgreater 1000 \textcolor{pp}{\textbf{GROUP BY}} T2.pPos}}\\
   \midrule
   \makecell{\textbf{Inference}} & 
   \makecell[l]{ 
    \textbf{\#\#\#\#} Please follow the previous example and help me generate the following SQL statement: \\
    Schema: ...\\
    Question: Return the names of conductors that do not have the nationality ``USA''.\\
    \ul{The target SQL only uses the following SQL keywords: WHERE; The difficulty rating of the target SQL is 100;
    }\\
    \textbf{\#\#\#\#} The target SQL is:
    }\\
    \bottomrule
  \end{tabularx}
  \label{tab:prompt}
\end{table*}

\subsubsection{Training Settings} The subsequent section explains the implementation specifics of the three models used in \textsc{Metasql}, i.e., the multi-label classification model, the first-stage ranking model, and the second-stage ranking model.


\noindent\textbf{Multi-label Classification Model.} As mentioned in Section \ref{section 3.1}, the multi-label classification model can be obtained from any NL2SQL translation model by substituting its top layer with a classification layer. In our experiments, we use \textsc{Lgesql} as the base translation model to implement.

\noindent\textbf{First-stage Ranking Model.} The embedding layer is initialized with publicly available pre-trained sentence-transformers \textsc{Stsb-mpnet-base-v2}\footnote{\href{https://huggingface.co/sentence-transformers/stsb-mpnet-base-v2}{https://huggingface.co/sentence-transformers/stsb-mpnet-base-v2}} model. In training, we use Adam \cite{Kingma14} optimizer with a learning rate of 2e-5 and warm up over the first 10\% of total steps. The batch size is set to 8. 


\noindent\textbf{Second-stage Ranking Model.} The model is based on \textsc{RoBERTa-large} \cite{RoBERTa19}. We use Adam optimizer with a learning rate of 1e-5 and adopt a schedule that reduces the learning rate by a factor of 0.5 once learning stagnates.

To further facilitate the listwise approach (as described in Section \ref{re-ranking model}), we configured the threshold $L$ to 10, which enabled us to generate a list of 10 SQL queries for each NL query. In addition, we set the batch size to $2$ per GPU in the training phase to expedite the process. Consequently, $60$ NL-SQL pairs were utilized in each training iteration.


\subsubsection{Inference Settings} 
Regarding to the multi-label classification model, we designated the classification threshold $p$ to be $0$, thereby enabling the selection of all conceivable query metadata labels. With respect to the first-stage ranking model, we configure it to select the top ten most highly ranked subsets from candidate SQL queries before passing the selected subset to the second-stage ranking model for final inference.


\subsubsection{Evaluation Metrics}
We adopt \textit{translation accuracy (EM)}, \textit{execution match (EX)}, \textit{translation precision}, and ranking metric \textit{translation MRR} \cite{MRR99} to assess model performance.

\noindent\textbf{Translation Accuracy} evaluates whether the top-1 generated SQL query matches the ``gold'' SQL; if it does, the translation is considered accurate. Otherwise, it is deemed inaccurate. It is a performance lower bound since a semantically correct SQL query may differ from the ``gold'' SQL query syntactically.

The metric is equivalent to the \textit{Exact Match Accuracy} metric proposed by \textsc{Spider}. It involves comparing sets for each SQL statement, and specific values are disregarded when conducting the accuracy calculation between the two SQL queries.

\noindent\textbf{Execution Accuracy} evaluates if the execution result matches the ground truth by executing the generated SQL query against the underlying relational database. This metric is the same as the \textit{Execution Match Accuracy} metric introduced in \textsc{Spider}.



\noindent \textbf{Translation Precision} at $K$ (denoted Precision@$K$) is the number of NL queries that an NLIDB system has the ``gold'' SQL queries in the top-$K$ translation results divided by the total number of NL queries. In our experiments, we choose $K$ to $1$, $3$, and $5$ to evaluate the performance of \textsc{Metasql}.


\definecolor{darkred}{rgb}{0.70, 0, 0}
\definecolor{Gray}{gray}{0.9}
\begin{table*}[ht]
  \captionsetup{font=small}
  \caption{Translation results on the two public NLIDB benchmarks.}
  \vspace{-2mm}
  
  \begin{adjustbox}{width=\textwidth}
  \begin{threeparttable}
  \begin{tabular}{l c c c c c c c } \hline
     & \multicolumn{2}{c}{\textbf{$\textsc{Spider}_{Dev}$}}  & \multicolumn{2}{c}{\textbf{$\textsc{Spider}_{Test}$}} & \multicolumn{3}{c}{\textbf{$\textsc{ScienceBenchmark}$}\tnote{1}} \\ 
     \cmidrule{2-8}
    \textbf{NLIDB Models}  & \textbf{EM\%} & \textbf{EX\%} & \textbf{EM\%} & \textbf{EX\%} & \textbf{EM\%}(\textsc{OncoMx}) & \textbf{EM\%}(\textsc{Cordis}) & \textbf{EM\%}(\textsc{Sdss}) \\
    \hline\hline
    \textsc{Bridge} \cite{BRIDGE20}  & 68.7 & 68.0 & 65.0 & 64.3 & 16.5 & 23.0 & 5.0 \\
    \rowcolor{Gray} \textsc{Bridge}+\textsc{Metasql}   & $70.5_{(\textcolor{darkred}{\uparrow 1.8})}$ & $69.2_{(\textcolor{darkred}{\uparrow 1.2})}$ & - & - & $\textbf{18.6}_{(\textcolor{darkred}{\uparrow\textbf{2.1}})}$ & $\textbf{25.0}_{(\textcolor{darkred}{\uparrow\textbf{2.0}})}$ & $\textbf{7.0}_{(\textcolor{darkred}{\uparrow\textbf{2.0}})}$ \\ \hline
    
    \textsc{Gap} \cite{GAP21}   & 71.8 & 34.9 & 69.7 & - & 33.0 & 20.0 & 5.0 \\ 
    \rowcolor{Gray} \textsc{Gap}+\textsc{Metasql}  & $73.4_{(\textcolor{darkred}{\uparrow 1.6})}$ & $37.2_{(\textcolor{darkred}{\uparrow 2.3})}$ & - & - & $\textbf{35.0}_{(\textcolor{darkred}{\uparrow\textbf{2.0}})}$ & 20.0 & $\textbf{6.0}_{(\textcolor{darkred}{\uparrow\textbf{1.0}})}$ \\ \hline
    
    \textsc{Lgesql} \cite{LGESQL20} & 75.1 & 36.3 & 72.0 & 34.2 & 41.7 & 24.0 & 4.0 \\
    \rowcolor{Gray} \textsc{Lgesql}+\textsc{Metasql} & $\textbf{77.4}_{(\textcolor{darkred}{\uparrow\textbf{2.3}})}$ & $\textbf{42.0}_{(\textcolor{darkred}{\uparrow\textbf{5.7}})}$ & $\textbf{72.3}_{(\textcolor{darkred}{\uparrow\textbf{0.3}})}$  & $\textbf{55.7}_{(\textcolor{darkred}{\uparrow\textbf{21.5}})}$ & $\textbf{42.7}_{(\textcolor{darkred}{\uparrow\textbf{1.0}})}$ & $\textbf{28.0}_{(\textcolor{darkred}{\uparrow\textbf{4.0}})}$ & $\textbf{12.0}_{((\textcolor{darkred}{\uparrow\textbf{8.0}})}$ \\\hline 
    
    $\textsc{Resdsql}_{\textsc{Large}}$ \cite{resdsql23}   & 75.8 & 80.1 & - & - & 42.7 & 29.0 & 4.0 \\ 
    \rowcolor{Gray} $\textsc{Resdsql}_{\textsc{Large}}$+\textsc{Metasql}  & $76.9_{(\textcolor{darkred}{\uparrow 1.1})}$ & $81.5_{(\textcolor{darkred}{\uparrow 1.4})}$ & - & - & $\textbf{49.7}_{(\textcolor{darkred}{\uparrow\textbf{7.0}})}$ & $\textbf{33.0}_{(\textcolor{darkred}{\uparrow\textbf{4.0}})}$ & $\textbf{10.0}_{(\textcolor{darkred}{\uparrow\textbf{6.0}})}$ \\ \Xhline{3\arrayrulewidth}
    
    \textsc{ChatGPT} & 51.5 & 65.3 & - & - & 51.2 & 40.0 & 11.0 \\
    \rowcolor{Gray} \textsc{ChatGPT}+\textsc{Metasql} & $\textbf{65.1}_{(\textcolor{darkred}{\uparrow\textbf{13.6}})}$ & $\textbf{74.2}_{(\textcolor{darkred}{\uparrow\textbf{8.9}})}$ & -  & - & $\textbf{53.2}_{(\textcolor{darkred}{\uparrow\textbf{2.0}})}$ & $\textbf{42.0}_{(\textcolor{darkred}{\uparrow\textbf{2.0}})}$ & $\textbf{16.0}_{(\textcolor{darkred}{\uparrow\textbf{5.0}})}$  \\\hline
    
    \textsc{Gpt-4} & 54.3 & 67.4 & - & - & 65.7 & 42.0 & 15.0 \\
    \rowcolor{Gray} \textsc{Gpt-4}+\textsc{Metasql} & $\textbf{69.6}_{(\textcolor{darkred}{\uparrow\textbf{15.3}})}$ & $\textbf{76.8}_{(\textcolor{darkred}{\uparrow\textbf{9.4}})}$ & -  & - & $\textbf{68.6}_{(\textcolor{darkred}{\uparrow\textbf{2.9}})}$& 42.0 & $\textbf{17.6}_{(\textcolor{darkred}{\uparrow\textbf{2.6}})}$ \\\hline
    
  \end{tabular}
  \begin{tablenotes}
  \item[1] As the database files for \textsc{Cordis} and \textsc{Sdss} are inaccessible, our evaluation is limited to the translation accuracy metric for \textsc{ScienceBenchmark}.
  \end{tablenotes}
  \end{threeparttable}
  \end{adjustbox}
  \label{tab:metasql overall accuracy results}
\end{table*}

\noindent\textbf{Translation MRR (Mean Reciprocal Rank)} is a statistic measure for evaluating an NLIDB system that provides a ranked list of SQL queries in response to each NL query. The metric is defined in the following way,

\begin{equation}
\resizebox{.45\hsize}{!}{$
MRR = \frac{1}{N} \sum^{N}_{i=1}\frac{1}{rank_{i}}
$}
\end{equation}


\noindent where $N$ denotes the number of given NL queries and $rank_{i}$ refers to the rank position of the ``gold'' SQL query for the $i^{th}$ NL query. Thus, the closer the value of MRR is to $1$, the more effective the translation ranking scheme is.

  

\subsection{Experimental Results}
We utilized \textsc{Metasql} with four Seq2seq NL2SQL translation models: \textsc{Bridge}, \textsc{Gap}, \textsc{Lgesql} and \textsc{Resdsql}\footnote{\textsc{Resdsql} model was implemented using three different scales of T5, namely Base, Large and 3B. We apply \textsc{Metasql} to \textsc{Resdsql} model with \textsc{T5-Large} scale, referred to as $\textsc{Resdsql}_{\textsc{Large}}$ in the following.}, in addition to two widely known LLMs, \textsc{ChatGPT} and \textsc{Gpt-4}. To evaluate the LLMs, we conduct experiments using the \textit{few-shot prompting} structure introduced in \cite{SQL-PaLM23}. This prompt structure entails providing instructions preceded by a few demonstrations (inputs, SQL) pairs\footnote{In our experiments, we use \textit{nine} demonstrations for each query.}, where the inputs are carefully crafted to include an NL question, a descriptive text about the database schema, including tables and columns, primary-foreign key specifications (optional)\footnote{We observed that since most column names in \textsc{Spider}'s databases are descriptive, LLMs can infer key relationships without explicit prompts. However, given that column names in \textsc{ScienceBenchmark}'s databases are mostly symbolic, it is essential to include key specifications in the prompt.}, along with supplementary metadata information for use with \textsc{Metasql}. (For specific details, refer to Table \ref{tab:prompt}.) We leverage the Whisper API\footnote{\url{https://openai.com/blog/introducing-chatgpt-and-whisper-apis}} provided by OpenAI to make the inference. 


 

Table \ref{tab:metasql overall accuracy results} summarizes the overall accuracy of the models. Overall, modern NL2SQL models demonstrate much better performance on \textsc{Spider} compared to \textsc{ScienceBenchmark}. In particular, due to the complexity of queries in the \textsc{Sdss} database of \textsc{ScienceBenchmark} (involving numerous {\fontfamily{qcr}\selectfont\textcolor{pp}{\textbf{WHERE}}} conditions and {\fontfamily{qcr}\selectfont\textcolor{pp}{\textbf{JOIN}}} operations), all models exhibit poor performance, hovering around 10\%. This underscores a notable challenge in handling queries in real-world databases.

\noindent\textbf{\textsc{Metasql} with Seq2seq Models.} The overall performance of all four baseline translation models can be consistently improved using \textsc{Metasql} across two benchmarks, with more noticeable improvements observed on \textsc{ScienceBenchmark} benchmark. It is worth noting that, except for \textsc{Bridge} and $\textsc{Resdsql}_{\textsc{Large}}$, the other two models (\textsc{Gap} and \textsc{Lgesql}) lack explicit handling of specific values in SQL queries. Consequently, the two models tend to exhibit lower execution accuracy compared to their translation accuracy on \textsc{Spider}.

One remarkable outcome is observed when applying \textsc{Metasql} to \textsc{Lgesql}. It attains an impressive 8.0\% improvement (from 4.0\% to 12.0\%) on the challenging \textsc{Sdss} database of \textsc{ScienceBenchmark}. Simultaneously, it achieves a translation accuracy of 77.4\% on the validation set and 72.3\% on the test set of the \textsc{Spider} benchmark, which is on par or higher than those of leading models on the \textsc{Spider} leaderboard. In addition, while \textsc{Lgesql} is not designed for value prediction, utilizing \textsc{Metasql} can significantly improve the execution accuracy by 5.7\% (and 21.5\%) on the validation and test set, respectively. The reason for this improvement is the explicit addition of values before the ranking procedure in \textsc{Metasql}.

\noindent\textbf{\textsc{Metasql with LLMs.}} Notably, \textsc{Metasql} significantly elevates the performance of \textsc{ChatGPT} and \textsc{Gpt-4}, compared to those Seq2seq-based counterparts. This substantial difference in improvement can be attributed to two key factors: 1) Given that \textsc{Metasql} relies on the underlying translation model to generate SQL candidates, the overall improvement largely depends on the quality of the SQL generation. Thanks to their powerful generation capability, modern LLMs can effectively harness \textsc{Metasql} to produce high-quality SQL candidates, yielding superior outputs. 2) LLMs serve as NL2SQL models without specific fine-tuning over existing benchmarks. This inherent diversity in the generation is complemented by the guidance from \textsc{Metasql}, enabling LLMs to align more effectively with benchmark-specific targeted outputs.

An outstanding result emerges from \textsc{Metasql} with \textsc{Gpt-4}, yielding a translation accuracy of 69.6\% and an execution accuracy of 76.8\% on \textsc{Spider} validation set, surpassing its performance by 15.3\% and 9.4\%, respectively. Furthermore, \textsc{Metasql} with \textsc{Gpt-4} attains a translation accuracy of 68.6\% on the \textsc{OncoMX} database of \textsc{ScienceBenchmark}.

\begin{table}[ht]
  \centering
  \captionsetup{font=small}
  \caption{EM(\%) on \textsc{Spider} validation set by SQL difficulty levels}
  \vspace{-2mm}
  \begin{adjustbox}{width=\linewidth}
  \begin{tabular}{l c c c c c} \hline
    \textbf{NL2SQL Models}  & \textbf{Easy}  & \textbf{Medium} & \textbf{Hard}  & \textbf{Extra Hard} & \textbf{Overall} \\ \hline\hline
    \textsc{Bridge} & 91.1 & 73.3  & 54.0 & 39.2     & 68.7   \\
    \rowcolor{Gray}\textsc{Bridge}+\textsc{Metasql} & $89.1_{(\downarrow 2.0)}$ & $\textbf{75.3}_{(\uparrow \textbf{2.0})}$  & $\textbf{58.0}_{(\uparrow \textbf{4.0})}$ & $42.8_{(\uparrow 3.6)}$  & 70.5   \\\hline
    
    \textsc{Gap}  & 91.5 & 74.2  & 64.4 & 44.2     & 71.8  \\ 
    \rowcolor{Gray}\textsc{Gap}+\textsc{Metasql} & $91.1_{(\downarrow 0.4)}$ & $\textbf{78.0}_{(\uparrow \textbf{3.8})}$  & $\textbf{64.9}_{(\uparrow \textbf{0.5})}$ & $43.4_{(\downarrow 0.8)}$  & 73.4   \\\hline
    
    \textsc{Lgesql}  & 91.9 & 77.4  & 65.5 & 53.0     & 75.1   \\
    \rowcolor{Gray}\textsc{Lgesql}+\textsc{Metasql} & $94.0_{(\uparrow 2.1)}$ & $\textbf{81.4}_{(\uparrow \textbf{4.0})}$  & $\textbf{70.1}_{(\uparrow \textbf{4.6})}$ & $49.4_{(\downarrow 3.6)}$     & 77.4   \\\hline
    
    $\textsc{Resdsql}_{\textsc{Large}}$ & 90.3 & 82.7  & 62.6 & 47.0     & 75.8   \\ \rowcolor{Gray}$\textsc{Resdsql}_{\textsc{Large}}$+\textsc{Metasql} & $92.5_{(\uparrow 2.2)}$ & $\textbf{83.9}_{(\uparrow \textbf{1.2})}$  & $\textbf{64.1}_{(\uparrow \textbf{1.5})}$ & $48.2_{(\uparrow 1.2)}$     & 76.9   \\ \Xhline{3\arrayrulewidth}
    
    \textsc{Chatpgt}  & 84.7 & 51.3  & 39.7 & 15.1     & 51.5  \\
    \rowcolor{Gray}\textsc{Chatpgt}+\textsc{Metasql} & $89.0_{(\uparrow 3.3)}$ & $\textbf{70.6}_{(\uparrow \textbf{19.3})}$  & $\textbf{55.2}_{(\uparrow \textbf{15.5})}$ & $24.4_{(\uparrow 9.3)}$     & 65.1   \\\hline
    
    \textsc{Gpt-4} & 82.2 & 56.3 & 51.3 & 14.6 & 54.3 \\
    \rowcolor{Gray} \textsc{Gpt-4}+\textsc{Metasql} & $91.1_{(\uparrow 8.9})$ & $\textbf{74.7}_{(\uparrow\textbf{18.4}})$ & $\textbf{64.1}_{(\uparrow\textbf{12.8}})$  & $36.1_{(\uparrow 21.5})$ & 69.6  \\\hline
    
  \end{tabular}
  \end{adjustbox}
  \label{tab:spider results by difficulty}
\end{table}

Next, we performed detailed experiments on \textsc{Spider} for \textsc{Metasql}. Table \ref{tab:spider results by difficulty} provides a breakdown of the translation accuracy on the \textsc{Spider} benchmark, categorized by the defined SQL difficulty levels. As expected, the performance of all the models drops with increasing difficulty. By applying \textsc{Metasql}, significant improvements are consistently observed for all translation models in the ``Medium'' and ``Hard'' queries, albeit with some degree of instability in other difficulty levels. For the instability observed in the ``Easy'' queries with \textsc{Bridge} and \textsc{Gap}, we find that \textsc{Metasql} occasionally ranks semantic-equivalent queries, leading to evaluation failures on the translation accuracy metric. On the other hand, for the instability of the ``Extra Hard'' queries with \textsc{Gap} and \textsc{Lgesql}, we attribute it primarily to inaccurate multi-grained signals that may be produced within complex queries, resulting in the incorrect ranking outcomes. This inaccuracy stems from the limitations of the rule-based approach outlined in Section \ref{reranking section}, where the pre-defined set of rules may fall short in addressing a SQL unit if its complexity is not explicitly considered (e.g., a nested query with more than two predicates).

We also present the accuracy results of \textsc{Metasql} compared with base models regarding SQL statement types in Table \ref{tab: results by syntax}. While the overall performance of six translation models can be effectively improved using \textsc{Metasql}, the breakdown results vary. Two findings from the results: (1) \textsc{Metasql} can significantly enhance query translations involving {\fontfamily{qcr}\selectfont \textcolor{pp}{\textbf{ORDER BY}}} and {\fontfamily{qcr}\selectfont\textcolor{pp}{\textbf{GROUP BY}}}-clauses, which is mainly due to the benefits derived from the ranking procedure. (2) Seq2seq-based translation models with \textsc{Metasql} deteriorate on translating nested-type complex queries (including {\fontfamily{qcr}\selectfont\textcolor{pp}{\textbf{NOT IN}}}-type negative queries), which the reason aligns with the instability observed in the ``Extra Hard'' queries discussed above.

\begin{table}[ht]
  \centering
  \aboverulesep=0ex
  \belowrulesep=0ex
  \captionsetup{font=small}
  \caption{EM(\%) on \textsc{Spider} validation set by SQL statement types}
  \vspace{-2mm}
  \begin{adjustbox}{width=\linewidth}
  \begin{tabular}{l c c c c c} \hline
    \textbf{NL2SQL Models}  & \textbf{Nested} & \textbf{Negation}  & \textbf{ORDERBY} & \textbf{GROUPBY} \\ \hline\hline
    
    \textsc{Bridge} & 42.8 & 52.9  & 63.6 & 56.8 \\ 
    \rowcolor{Gray}\textsc{Bridge}+\textsc{Metasql} & $39.6_{(\downarrow 3.2)}$ & $49.5_{(\downarrow 3.4)}$  & $\textbf{70.6}_{(\uparrow\textbf{7.0})}$ & $\textbf{63.8}_{(\uparrow\textbf{7.0})}$ \\\hline
    
    \textsc{Gap}  & 47.2 & 60.0  & 71.0 & 67.9   \\  
    \rowcolor{Gray}\textsc{Gap}+\textsc{Metasql} & $44.7_{(\downarrow 2.5)}$ & $56.8_{(\downarrow 3.2})$  & $\textbf{73.2}_{(\uparrow\textbf{1.8})}$ & $\textbf{68.6}_{(\uparrow\textbf{0.7})}$  \\\hline
    
    \textsc{Lgesql}  & 54.1 & 62.1  & 74.9 & 67.9    \\ 
    \rowcolor{Gray}\textsc{Lgesql}+\textsc{Metasql} & $51.6_{(\downarrow 2.5)}$ & $62.1_{(-)}$  & $\textbf{78.8}_{(\uparrow\textbf{3.9})}$ & $\textbf{69.7}_{(\uparrow\textbf{1.8})}$ \\ \hline
    
     $\textsc{Resdsql}_{\textsc{Large}}$  & 50.3 & 57.9  & 74.0 & 72.0     \\      \rowcolor{Gray}$\textsc{Resdsql}_{\textsc{Large}}$+\textsc{Metasql} & $50.0_{(\downarrow 0.3)}$ & $59.1_{(\uparrow 1.2})$  & $\textbf{75.6}_{(\uparrow\textbf{1.6})}$ & $\textbf{73.1}_{(\uparrow\textbf{1.1})}$ \\\Xhline{3\arrayrulewidth}
     
    \textsc{ChatGPT}  & 28.3 & 47.4  & 42.0 & 29.5   \\ 
    \rowcolor{Gray}\textsc{ChatGPT}+\textsc{Metasql} & $43.1_{(\uparrow 14.8)}$ & $50.7_{(\uparrow 13.3)}$  & $\textbf{54.5}_{(\uparrow\textbf{12.5})}$ & $\textbf{44.4}_{(\uparrow\textbf{14.9})}$ \\\hline
    
    \textsc{Gpt-4} & 33.3 & 45.0 & 46.0 & 36.5 \\
    \rowcolor{Gray} \textsc{Gpt-4}+\textsc{Metasql} & $47.2_{(\uparrow\textbf{13.9}})$ & $55.0_{(\uparrow\textbf{10.0}})$ & $\textbf{74.0}_{(\uparrow\textbf{28.0}})$  & $\textbf{51.9}_{(\uparrow\textbf{15.4}})$ \\\hline
  \end{tabular}
  \end{adjustbox}
  \label{tab: results by syntax}
\end{table}

To assess the performance of the ranking pipeline in \textsc{Metasql}, Table \ref{tab:ranking result} shows the translation precision and MRR results on \textsc{Spider} validation set. Note that MRR values are calculated treating the reciprocal rank as $0$ when the ``gold'' query is not among the final top-5 results for a given NL query.

As can be seen in Table \ref{tab:ranking result}, \textsc{Metasql} with $\textsc{Resdsql}_{\textsc{Large}}$ attains a translation MRR of 78.8\%, surpassing the other models. The results also demonstrate that \textsc{Metasql} can correctly select the target SQL queries in the first few returned ranking results in most cases. This compelling evidence highlights its effectiveness, especially when compared with existing auto-regressive decoding techniques utilizing beam search or sampling methods. In particular, \textsc{Metasql} with \textsc{Lgesql} (and with \textsc{Gap}) achieves about 81.0\% translation precision in the top-5 retrieved results. 

\begin{table}[ht]
  \centering
  \captionsetup{font=small}
  \caption{Precision and MRR (\%) on \textsc{Spider} validation set}
  \vspace{-2mm}
  \begin{adjustbox}{width=\linewidth}
  \begin{tabular}{l c c c c} \hline
    \textbf{NL2SQL Models}  & \textbf{MRR} & \textbf{Precision@1}  & \textbf{Precision@3}  & \textbf{Precision@5}  \\ \hline\hline
    \textsc{Bridge}+\textsc{Metasql}   & 73.8 & 70.5 & 76.7 & 78.6 \\
    \textsc{Gap}+\textsc{Metasql}   & 76.4 & 73.4 & 79.9 & \textbf{81.0} \\
    \textsc{Lgesql}+\textsc{Metasql}   & 78.2 & 76.8 & 79.6 & \textbf{80.9} \\
    $\textsc{Resdsql}_{\textsc{Large}}$+\textsc{Metasql}   & 78.8 & 77.2 & 80.6 & \textbf{80.1} \\
    \textsc{ChatGPT}+\textsc{Metasql}   & 52.6 & 51.5 & 64.3 & 64.5 
    \\
    \textsc{Gpt-4}+\textsc{Metasql}   & 69.6 & 69.6 & 72.5 & 72.5 \\
    \hline
  \end{tabular}
  \end{adjustbox}
  \label{tab:ranking result}
\end{table}

Concerning a multi-stage solution like \textsc{Metasql}, a natural question arises about the potential impact of different stages on the overall outcome. To deepen our understanding of \textsc{Metasql}, we experimented on \textsc{Spider} validation set to evaluate the performance of each stage: For the first stage (i.e., metadata selection), we evaluated the accuracy of this stage by checking if predicted metadata labels could compose the ground-truth metadata. The second stage (i.e., metadata-conditioned generation) accuracy was determined by assessing if generated SQL queries, conditioned on metadata compositions from ground-truth labels, matched the ``gold'' query. For the last ranking stage, the accuracy was evaluated using the translation MRR, where the NL2SQL model generated ranking candidates conditioned on metadata compositions from ground-truth labels. Table~\ref{tab:bucket effect result} presents the accuracy results.

\begin{table}[ht]
  \centering
  \captionsetup{font=small}
  \caption{Stage-wise accuracy (\%), with bracketed values indicating the performance of the respective base models. As we utilized a unified multi-label classifier (implemented based on \textsc{Lgesql}) in our experiments, the accuracy remains consistent in the first stage.}
  \vspace{-2mm}
  \begin{adjustbox}{width=\linewidth}
  \begin{tabular}{c c c c } \hline
    \textbf{Model}  & \textbf{\makecell{Metadata Selection \\ Accuracy}} & \textbf{\makecell{Metadata-conditioned \\ Generation Accuracy}} & \textbf{\makecell{Ranking \\ Accuracy}}   \\ \hline
    \textsc{Bridge}+\textsc{Metasql}  & 91.4 & 77.3 (68.7) & 87.1 \\
    \textsc{Gap}+\textsc{Metasql}  & 91.4 & 77.9 (71.8) & 88.4 \\
    \textsc{Lgesql}+\textsc{Metasql} & 91.4 & 82.7 (75.1) & 90.3 \\
    $\textsc{Resdsql}_{\textsc{Large}}$+\textsc{Metasql} & 91.4 & 83.1 (75.8) & 89.6  \\\hline
  \end{tabular}
  \end{adjustbox}
  \label{tab:bucket effect result}
\end{table}

As can be seen, the performance of each stage remains consistently accurate across all three stages, while the second stage exhibits relatively notable performance fluctuations, attributed to the inherent limitations in underlying translation models. The results illustrate that \textsc{Metasql} effectively optimizes the performance of each stage in the current settings, thereby contributing to overall performance improvements.

\subsection{Metadata Sensitivity Analysis}






\pgfdeclareplotmark{mystar}{
    \node[star,star point ratio=2.25,minimum size=11pt, inner sep=0pt, draw=none, solid, fill=sea] {};
}
\begin{figure*}
  \begin{subfigure}{.47\textwidth}
    \centering
    \begin{tikzpicture}
    \begin{axis}[
        height=5.0cm,
        width=\linewidth,
        xlabel={Classification Threshold $p$},
        ylabel={Translation Accuracy},
        y label style={at={(axis description cs:-0.10,.5)},anchor=south},
        axis line style=ultra thick,
        xmin=-65, xmax=5,
        ymin=0.6, ymax=0.85,
        xtick={-60,-40,-20,-10,-5,0},
        ytick={0.7,0.751,0.774,0.8},
        legend pos=south east,
        legend style={
          nodes={scale=0.7, transform shape},
          very thin  
        },
        ymajorgrids=true,
        xmajorgrids=true,
        grid style=dashed,
        every axis plot/.append style={ultra thick}
    ]
    \addplot[
        sea,
        mark=mystar,
    ]
    coordinates {
        (0,0.774)(-5, 0.763)(-10,0.754)(-20,0.750)(-40,0.740)(-60,0.685)
    };
    \addplot[
        dashed,
        dustyorange,
    ]
    coordinates {
        (-60,0.751)(-40,0.751)(-20,0.751)(-10,0.751)(-5,0.751)(0,0.751)
    };
    \legend{\textsc{Lgesql}+$\textsc{Metasql}_{p}$, \textsc{Lgesql}}
    \end{axis}
    \end{tikzpicture}
    \captionsetup{font=small}
    \caption{Translation accuracy of \textsc{Metasql} with different classification thresholds. The dotted line shows the results of original \textsc{Lgesql}.}
    \label{fig: metasql multi-label results}
    \end{subfigure}\hspace{2em}
  \begin{subfigure}{.47\textwidth}
    \centering
    \begin{tikzpicture}
    \begin{axis}[
        height=5.0cm,
        width=\linewidth,
        xlabel={Classification Threshold $p$},
        ylabel={Translation Accuracy},
        y label style={at={(axis description cs:-0.09,.5)},anchor=south},
        axis line style=ultra thick,
        xmin=-65, xmax=5,
        ymin=0.5, ymax=0.85,
        xtick={-60,-40,-20,-10,-5,0},
        ytick={0.6,0.7,0.8},
        legend pos=south east,
        legend style={
          nodes={scale=0.7, transform shape},
          very thin  
        },
        xmajorgrids=true,
        ymajorgrids=true,
        grid style=dashed,
        every axis plot/.append style={ultra thick}
    ]
    \addplot[
        sea,
        mark=mystar,
    ]
    coordinates {
        (0,0.774)(-5, 0.763)(-10,0.754)(-20,0.750)(-40,0.740)(-60,0.685)
    };
    \addplot[
        indigo,
        mark=*,
        mark size=3pt,
    ]
    coordinates {
        (0,0.735)(-5,0.723)(-10,0.715)(-20,0.713)(-40,0.711)(-60,0.640)
    };
    \addplot[
        dustyorange,
        mark=triangle*,
        mark size=3pt,
        domain=-60:0
    ]
    coordinates {
        (0,0.744)(-5,0.741)(-10,0.731)(-20,0.719)(-40,0.70)(-60,0.659)
    };
    \legend{Correct Correctness Indicator, Incorrect Correctness Indicator, No Correctness Indicator}
    \end{axis}
    \end{tikzpicture}
    \captionsetup{font=small}
    \caption{Translation accuracy of \textsc{Metasql} applied to \textsc{Lgesql} using different correct indicator-type of metadata information.}
    \label{fig:correct indicator results}
    \end{subfigure}
  \begin{subfigure}{.47\textwidth}
    \centering
    \begin{tikzpicture}
    \begin{axis}[
        height=5.0cm,
        width=\linewidth,
        xlabel={Hardness Value},
        ylabel={Translation Accuracy},
        axis line style=ultra thick,
        xmin=50, xmax=950,
        ymin=0.6, ymax=0.85,
        xtick={100,300,500,700,900},
        ytick={0.7,0.768,0.8},
        legend pos=south east,
        legend style={
          nodes={scale=0.7, transform shape},
          very thin  
        },
        ymajorgrids=true,
        xmajorgrids=true,
        grid style=dashed,
        every axis plot/.append style={ultra thick}
    ]
    \addplot[
        sea,
        mark=mystar,
    ]
    coordinates {
        (100,0.759)(300,0.754)(500,0.750)(700,0.748)(900,0.745)
    };
    \addplot[
        dotted,
        color=indigo,
    ]
    coordinates {
        (100,0.780)(300,0.780)(500,0.780)(700,0.780)(900,0.780)
    };
    \addplot[
        dashed,
        color=dustyorange,
    ]
    coordinates {
        (100,0.774)(300,0.774)(500,0.774)(700,0.774)(900,0.774)
    };
    
    \legend{Fixed Hardness Values, Oracle Hardness Values, \textsc{Lgesql}+\textsc{Metasql}}
    \end{axis}
    \end{tikzpicture}
    \captionsetup{font=small}
    \caption{Translation accuracy of \textsc{Metasql} with different hardness values. The dotted line represents the result with the oracle setting.}
    \label{fig:metasql rating results}
    \end{subfigure}\hspace{3em}
  \begin{subfigure}{.47\textwidth}
    \centering
    \begin{tikzpicture}
    \begin{axis}[
        height=5.0cm,
        width=\linewidth,
        xlabel={Classification Threshold $p$},
        ylabel={Translation Accuracy},
        y label style={at={(axis description cs:-0.10,.5)},anchor=south},
        axis line style=ultra thick,
        xmin=-65, xmax=5,
        ymin=0.6, ymax=0.85,
        xtick={-60,-40,-20,-10,-5,0},
        ytick={0.7,0.8},
        legend pos=south east,
        legend style={
          nodes={scale=0.7, transform shape},
          very thin  
        },
        ymajorgrids=true,
        xmajorgrids=true,
        grid style=dashed,
        every axis plot/.append style={ultra thick}
    ]
    \addplot[
        dustyorange,
        mark=triangle*,
        mark size=3pt,
        domain=-60:0
    ]
    coordinates {
        (0,0.745)(-5,0.737)(-10,0.736)(-20,0.732)(-40,0.729)(-60,0.723)
    };
    \addplot[
        indigo,
        mark=*,
        mark size=3pt,
        domain=-60:0
    ]
    coordinates {
        (0,0.813)(-5,0.810)(-10,0.807)(-20,0.794)(-40,0.790)(-60,0.783)
    };
    \addplot[
        sea,
        mark=mystar,
    ]
    coordinates {
        (0,0.774)(-5, 0.763)(-10,0.754)(-20,0.750)(-40,0.740)(-60,0.685)
    };
    \legend{Randomized Operator Tags, Oracle Operator Tags, \textsc{Lgesql}+$\textsc{Metasql}_{p}$}
    \end{axis}
    \end{tikzpicture}
    \captionsetup{font=small}
    \caption{Translation accuracy of \textsc{Metasql} with different operator tags. The purple dotted line represents the result with the oracle setting.}
    \label{fig:metasql operator results}
    \end{subfigure}
    \vspace{-6mm}
    \captionsetup{font=small}
    \caption{Metadata sensitivity analysis on \textsc{Metasql}}
    \label{fig:sensitivity}

\end{figure*}
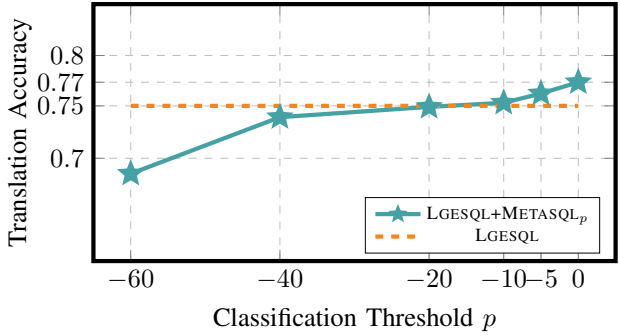
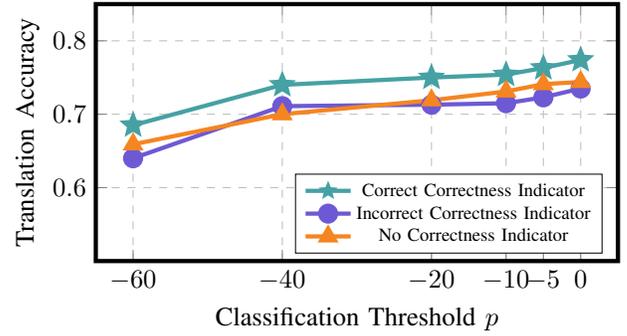
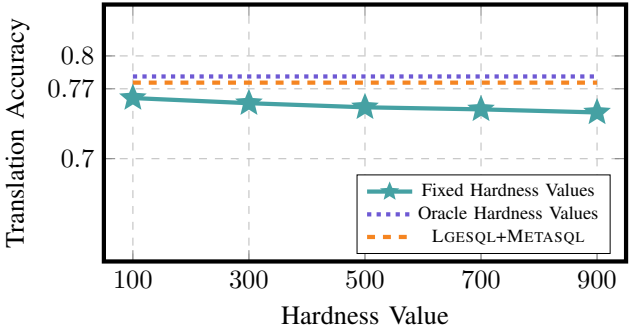
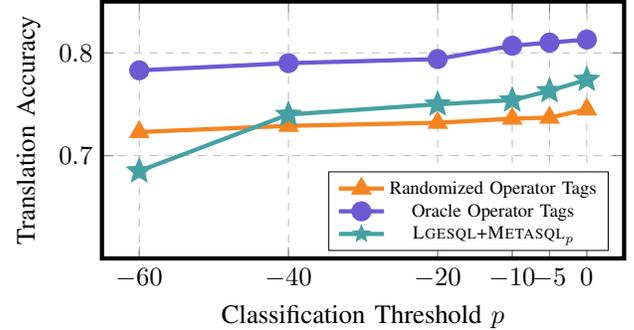

To gain a deeper understanding of \textsc{Metasql}, we perform a sensitivity analysis of the query metadata, specifically on \textsc{lgesql}. Our exploration revolves around two key questions: 1) \textit{Does the model respond appropriately to variations in this conditioning metadata?} and 2) \textit{What are the optimal settings for generating this metadata during testing when it is inaccessible?}  The experimental results are detailed in Fig.~\ref{fig:sensitivity}.


\noindent\textbf{Metadata Selection Rate} (Fig.~\ref{fig: metasql multi-label results}). To begin, we examine the sensitivity of \textsc{Metasql} concerning the metadata selected from the multi-label classifier in order to measure the importance of metadata quality to model performance. For this purpose, we intentionally select more ``noisy'' metadata by systematically reducing the classification threshold $p$ from its default value of $0$ to its minimum predicted value of $-60$, effectively leading to a ``randomized'' metadata selection scenario.

The findings indicate a strong dependence of \textsc{Metasql}'s performance on the metadata selected from the multi-label classifier.  \textbf{\textit{With more ``noisy'' metadata involved, the improvements yielded by \textsc{Metasql} diminish significantly, and in some cases, even lead to performance degradation.}} Particularly, since the multi-label classifier tends to generate high-confidence predictions, a significant performance drop is observed when $p$ is lower than $-10$, primarily due to the increased involvement of ``noisy'' metadata.


\noindent\textbf{Correctness Indicator} (Fig.~\ref{fig:correct indicator results}). Here, our focus lies in investigating the extent to which \textsc{Metasql} relies on this metadata by supplying either an incorrect or even no indicator.

Overall, \textsc{Metasql} responds appropriately to the changes of this metadata, experiencing a reduction in performance when conditioned on an incorrect indicator or no correctness indicator is provided. It is worth noting that conditioning on incorrect correctness indicator leads to slightly worse performance than the latter scenario, indicating that \textbf{\textit{providing incorrect metadata may have a more detrimental impact on \textsc{Metasql} compared to providing no metadata at all.}}



\noindent\textbf{Hardness Value} (Fig.~\ref{fig:metasql rating results}). We examine how the hardness values provided to \textsc{Metasql} affect its performance. Our experiment involves two configurations: 1) we maintain a fixed hardness value, independent of the query, and 2) we provide the oracle hardness values during the inference time.

The findings reveal that \textbf{\textit{the performance of \textsc{Metasql} remains relatively stable with changes in the hardness values.}} This is attributed to two aspects: 1) The hardness values obtained from the multi-label classifier closely align with the oracle values in most cases, and \textsc{Metasql} can generate correct SQL queries even when the inference-time hardness values are not identical to the oracle ones. 2) \textsc{Metasql} tends to incorporate various types of metadata globally, rather than relying solely on a specific type of metadata. In other words, compared to other types of metadata, \textsc{Metasql} shows lower sensitivity to this particular metadata.

Moreover, an intriguing finding is that specifying an easier hardness value tends to yield better results than a harder one, while still achieving worse performance than its current setting. We posit that this is because existing translation models often perform better on relatively straightforward queries.

\noindent\textbf{Operator Tags} (Fig.~\ref{fig:metasql operator results}). To assess the significance of operator tag-type metadata, we analyze model performance in response to various changes in this metadata. To experiment, we employ two distinct settings: (1) we provide the oracle set of operator tags for each query, and (2) we randomize operator tags.

The findings reveal that \textbf{\textit{\textsc{Metasql} exhibits greater sensitivity to operator tag-type metadata compared to other types of metadata.}} The reason we believe is that this metadata can provide useful generation constraints for \textsc{Metasql}, thereby reducing the search space for the underlying model during the auto-regressive decoding procedure, and hence resulting in improved outcomes. Notably, with the aid of the oracle operator tags, \textsc{Metasql} with default classification threshold setting ($p=0$) can attain a translation accuracy of 81.3\%.

\subsection{Ablation Study}

\begin{table}[ht]
  \centering
  \captionsetup{font=small}
  \caption{Ablation study on \textsc{Spider} validation set. The ``w/o Multi-label Classifier'' denotes candidate SQL queries generation with all metadata compositions and the ``w/o Phrase-level Supervision'' denotes removing NL-to-Phrase and Phrase Triplet loss from training.}
  \vspace{-2mm}
  \begin{adjustbox}{width=\linewidth}
  {\begin{tabular}{c c c c} \hline
    \textbf{Model}  & \textbf{\makecell{Generation \\ Miss Count}} & \textbf{\makecell{Ranking \\ Miss Count}} & \textbf{Overall(\%)}   \\ \hline\hline
    Base Model (\textsc{Lgesql} + \textsc{Metasql})  & 185 & 56 & 77.4 \\
    w/o Multi-label Classifier  & 167 & 159 & $68.5_{(\downarrow 8.8)}$ \\
    \;\;\;\;\; w/o Phrase-level Supervision & 185 & 87 & $75.2_{(\downarrow 2.2)}$ \\
    \;\;\;\;\;\;\;\;\;\;\; w/o Second-stage Ranking Model & 185 & 253 & $57.7_{(\downarrow\downarrow \textbf{19.7})}$ \\ 
    \hline
  \end{tabular}}
  \end{adjustbox}
  \label{tab:ablation study results}
\end{table}

We conduct an ablation study on \textsc{Spider} validation set with \textsc{Lgesql} to explore the efficacy of the multi-label classifier and the second-stage ranking model. Our experiment involves comparing three distinct settings: (1) brute-force generation of candidate SQL queries by utilizing all possible metadata, namely without a multi-label classifier, (2) controlling inclusion of fine-grained features in the second-stage ranking model and (3) employing the second-stage ranking model. 

The results are presented in Table \ref{tab:ablation study results}. The findings reveal that our ranking process experiences a significant decline in performance when it fails to capture relevant metadata or exclude the second-stage precise ranking model. (Despite some gains during the generation process in the former one.) Additionally, the results demonstrate the essential role of fine-grained supervision signal in the second-stage ranking model, as the performance experiences a notable drop without it, further emphasizing its significance in our approach.

\subsection{Analysis of \textsc{Metasql}}

To better understand \textsc{Metasql}, we analyzed the translation results on \textsc{Spider} validation set. We identify the following three major categories for the failures.



 \begin{itemize}[leftmargin=*]
 \item \textbf{Auto-regressive Decoding Problem.} A significant number of translation errors in the generation process can be attributed to the limitations of auto-regressive decoding used in existing translation models. This means that despite accurate metadata provided by \textsc{Metasql}, the underlying translation model may still produce incorrect translations. Such errors are particularly noticeable in some complex queries, as demonstrated by the following example, 

\vspace{1mm}
\noindent\textbf{NL Query:} \textit{What major is every student who does not own a cat as a pet, and also how old are they?}\\
\noindent\textbf{Gold SQL Query:} 

{\fontfamily{qcr}\fontsize{9}{10}\selectfont\textcolor{pp}{\textbf{SELECT}} major, age \textcolor{pp}{\textbf{FROM}} student} \\
{\fontfamily{qcr}\fontsize{9}{10}\selectfont \textcolor{pp}{\textbf{WHERE}} stuid \textcolor{pp}{\textbf{NOT IN}} (} \\
\hspace*{0.1cm}{\fontfamily{qcr}\fontsize{9}{10}\selectfont \textcolor{pp}{\textbf{SELECT}} T1.stuid \textcolor{pp}{\textbf{FROM}} student \textcolor{pp}{\textbf{AS}} T1} \\
\hspace*{0.1cm}{\fontfamily{qcr}\fontsize{9}{10}\selectfont \textcolor{pp}{\textbf{JOIN}} has\_pet \textcolor{pp}{\textbf{AS}} T2 \textcolor{pp}{\textbf{JOIN}} pets \textcolor{pp}{\textbf{AS}} T3} \\
\hspace*{0.1cm}{\fontfamily{qcr}\fontsize{9}{10}\selectfont \textcolor{pp}{\textbf{WHERE}} T3.pettype = \textcolor{rd}{\textquotesingle cat\textquotesingle})}

\noindent\textbf{Incorrect Generated SQL Query:}

{\fontfamily{qcr}\fontsize{9}{10}\selectfont \textcolor{pp}{\textbf{\textbf{SELECT}}} major, age \textcolor{pp}{\textbf{\textbf{FROM}}} student } \\
{\fontfamily{qcr}\fontsize{9}{10}\selectfont \textcolor{pp}{\textbf{\textbf{WHERE}}} stuid \textcolor{pp}{\textbf{\textbf{NOT IN}}} ( } \\
\hspace*{0.1cm}{\fontfamily{qcr}\fontsize{9}{10}\selectfont \textcolor{pp}{\textbf{\textbf{SELECT}}} has\_pet.stuid \textcolor{pp}{\textbf{\textbf{FROM}}} has\_pet \textcolor{pp}{\textbf{\textbf{JOIN}}} pets} \\
\hspace*{0.1cm}{\fontfamily{qcr}\fontsize{9}{10}\selectfont \textcolor{pp}{\textbf{\textbf{WHERE}}} pets.pettype = \textcolor{rd}{\textquotesingle cat\textquotesingle})}
\vspace{1mm}

 \noindent As can be seen, \textsc{Metasql} fails to generate the correct join path (i.e., {\fontfamily{qcr}\selectfont student-has\_pet-pets}) used in the nested query, even though the generation is conditioned on the oracle query metadata (i.e., \textit{450}, \textit{where}, \textit{subquery}). To a certain degree, enabling model sampling during query generation may mitigate such failures, but enhancing the performance of the translation model is crucial for long-term improvements in accuracy.
 
 \item \textbf{Metadata Mismatch Problem.} Another large portion of translation errors in the generation process is due to inaccurate query metadata retrieved from the multi-label classifier. For example, the following is an example in \textsc{Spider},

\vspace{1mm}
\noindent\textbf{NL Query:} \textit{How many countries has more than 2 makers?}\\
\noindent\textbf{Oracle Metadata:} \textit{200, group, join}\\
\noindent\textbf{Predicted Metadata:} \textit{350, group, subquery}\\
\noindent\textbf{Gold SQL Query:} 

{\fontfamily{qcr}\fontsize{9}{10}\selectfont \textcolor{pp}{\textbf{SELECT}} \textcolor{blue}{count}(*) \textcolor{pp}{\textbf{FROM}}} \\
{\fontfamily{qcr}\fontsize{9}{10}\selectfont countries \textcolor{pp}{\textbf{AS}} T1 \textcolor{pp}{\textbf{JOIN}} car\_makers \textcolor{pp}{\textbf{AS}} T2} \\
{\fontfamily{qcr}\fontsize{9}{10}\selectfont \textcolor{pp}{\textbf{GROUP BY}} T1.countryid \textcolor{pp}{\textbf{HAVING}} \textcolor{blue}{count}(*)>2}

\noindent\textbf{Incorrect Generated SQL Query:}

{\fontfamily{qcr}\fontsize{9}{10}\selectfont \textcolor{pp}{\textbf{SELECT}} \textcolor{blue}{count}(*) \textcolor{pp}{\textbf{FROM}} ( } \\
\hspace*{0.1cm}{\fontfamily{qcr}\fontsize{9}{10}\selectfont \textcolor{pp}{\textbf{SELECT}} country \textcolor{pp}{\textbf{FROM}} car\_makers} \\
\hspace*{0.1cm}{\fontfamily{qcr}\fontsize{9}{10}\selectfont \textcolor{pp}{\textbf{GROUP BY}} country \textcolor{pp}{\textbf{HAVING}} \textcolor{blue}{count}(*)>2)}
\vspace{1mm}

\noindent Given that \textsc{Metasql} erroneously extracts \textit{subquery} metadata, the underlying translation model was altered to generate a query resembling a subquery. As a result, it is imperative to establish a more dependable approach for selecting pertinent metadata for \textsc{Metasql}.

 \item \textbf{Ranking Problem.} Many mistranslations stem from the ranking procedure, primarily in the second stage. Even when the ``gold'' query is included as a candidate, \textsc{Metasql} may not prioritize the ``gold'' query at the top position. Such failures are commonly observed in queries with join operations, where the increased abstraction in query semantics poses challenges. An illustrative example is provided below,

\vspace{1mm}
\noindent\textbf{NL Query:} \textit{Which car models are produced after 1980?}\\
\noindent\textbf{Gold SQL Query:} 

{\fontfamily{qcr}\fontsize{9}{10}\selectfont \textcolor{pp}{\textbf{SELECT}} T1.model \textcolor{pp}{\textbf{FROM}} model\_list \textcolor{pp}{\textbf{AS}} T1 } \\
{\fontfamily{qcr}\fontsize{9}{10}\selectfont \textcolor{pp}{\textbf{JOIN}} car\_names \textcolor{pp}{\textbf{AS}} T \textcolor{pp}{\textbf{JOIN}} car\_data \textcolor{pp}{\textbf{AS}} T3} \\
{\fontfamily{qcr}\fontsize{9}{10}\selectfont \textcolor{pp}{\textbf{WHERE}} T3.year > \textcolor{rd}{1980}}

\noindent\textbf{Top-ranked SQL Query:}

{\fontfamily{qcr}\fontsize{9}{10}\selectfont \textcolor{pp}{\textbf{SELECT}} T2.model \textcolor{pp}{\textbf{FROM}} cars\_data \textcolor{pp}{\textbf{AS}} T1} \\
{\fontfamily{qcr}\fontsize{9}{10}\selectfont \textcolor{pp}{\textbf{JOIN}} car\_names \textcolor{pp}{\textbf{AS}} T2 \textcolor{pp}{\textbf{WHERE}} T1.year > \textcolor{rd}{1980}}
\vspace{1mm}

\noindent Such failures may be eliminated if more specific semantics over the underlying database can be captured and incorporated into the training of the ranking model. 
 \end{itemize}
 
From the above analysis, we enhance our comprehension of various aspects of \textsc{Metasql} and explore some improvements that can be made in the future.

\section{Related Work}\label{section 5}
NLIDBs have been studied for decades both in the database management and NLP communities. Early works \cite{Androutsopoulos95, GEO96, precise08, Zettlemoyer05, NaLR14, ATHENA16, TEMPLAR19} employ rule-based approaches with handcrafted grammars to map NL queries to database-specific SQL queries. The recent rise of deep learning leads to machine learning-based approaches, treating NLIDB as a Seq2seq translation task using the encoder-decoder architecture \cite{sqlnet17, Yu18, Bogin19, Guo19, Wang20, Yu20, BRIDGE20, SmBop20, LGESQL20, resdsql23, IKnowSQL23, CatSQL23, MIGA22}. However, these Seq2seq-based methods, due to their auto-regressive decoding nature, face limitations in handling complex queries. Instead of relying on standard auto-regressive decoding, \textsc{Metasql} uses control signals to better control SQL generation, resulting in improved outcomes.

With the excellent success of LLMs in various NLP tasks, recent works have explored applying LLMs to the NL2SQL task \cite{nitarshan22, aiwei23, DINSQL23, Interleaving23, SQL-PaLM23}. \cite{nitarshan22, aiwei23} systematically evaluate the NL2SQL capabilities of existing LLMs. To optimize the LLM prompting, recent studies \cite{SQL-PaLM23, DINSQL23} have curated detailed prompts for improved SQL query generation. Moreover, a more recent study \cite{Interleaving23} aims to capitalize on the complementary strengths of fine-tuned translation models and LLMs, striving for zero-shot NL2SQL support. Unlike various existing approaches, \textsc{Metasql} introduces a unified framework that harnesses the advantages of existing LLMs and further enhances their translation performance.

\section{Conclusion \& Future Work}\label{section 6}
This paper proposed a unified framework named \textsc{Metasql} for the NL2SQL problem, which can be used for any existing translation models to enhance their performance. Instead of parsing NL query into SQL query end to end, \textsc{Metasql} exploits the idea of controllable text generation by introducing query metadata for better SQL query candidates generation and then uses learning-to-rank algorithms to retrieve globally optimized queries. Experimental results showed that the performance of six translation models can be effectively enhanced after applying \textsc{Metasql}. Moreover, we conduct detailed analysis to explore various aspects of \textsc{Metasql}, which gain more insights on this novel generate-then-rank approach.

 Although \textsc{Metasql} has demonstrated its effectiveness in its current form, these results call for further future work in this direction. One potential area of investigation is how to extend the generate-then-rank approach beyond the existing auto-regressive decoding paradigm, allowing \textsc{Metasql} to overcome the limitations observed in the decoding procedure of existing translation models and hence further improve their performance. Additionally, developing a more precise multi-grained semantics labeling method, particularly for those complex queries, in the ranking process is critical for further enhancing the performance of \textsc{Metasql}. Finally, an intended future research direction is exploring the possibility of integrating other types of metadata into \textsc{Metasql}.



\bibliographystyle{IEEEtran}
\bibliography{IEEEexample}

\begin{thebibliography}{10}
\providecommand{\url}[1]{#1}
\csname url@samestyle\endcsname
\providecommand{\newblock}{\relax}
\providecommand{\bibinfo}[2]{#2}
\providecommand{\BIBentrySTDinterwordspacing}{\spaceskip=0pt\relax}
\providecommand{\BIBentryALTinterwordstretchfactor}{4}
\providecommand{\BIBentryALTinterwordspacing}{\spaceskip=\fontdimen2\font plus
\BIBentryALTinterwordstretchfactor\fontdimen3\font minus \fontdimen4\font\relax}
\providecommand{\BIBforeignlanguage}[2]{{%
\expandafter\ifx\csname l@#1\endcsname\relax
\typeout{** WARNING: IEEEtran.bst: No hyphenation pattern has been}%
\typeout{** loaded for the language `#1'. Using the pattern for}%
\typeout{** the default language instead.}%
\else
\language=\csname l@#1\endcsname
\fi
#2}}
\providecommand{\BIBdecl}{\relax}
\BIBdecl

\bibitem{Androutsopoulos95}
I.~Androutsopoulos, G.~D. Ritchie, and P.~Thanisch, ``Natural language interfaces to databases - an introduction,'' \emph{Nat. Lang. Eng.}, no.~1,  29--81, 1995.

\bibitem{Visionnary99}
F.~Benzi, D.~Maio, and S.~Rizzi, ``{VISIONARY:} a viewpoint-based visual language for querying relational databases,'' \emph{J. Vis. Lang. Comput.}, no.~2,  117--145, 1999.

\bibitem{Banks02}
G.~Bhalotia, A.~Hulgeri, C.~Nakhe, S.~Chakrabarti, and S.~Sudarshan, ``Keyword searching and browsing in databases using {BANKS},'' in \emph{ICDE}, R.~Agrawal and K.~R. Dittrich, Eds., 2002,  431--440.

\bibitem{sqlnet17}
X.~Xu, C.~Liu, and D.~Song, ``Sqlnet: Generating structured queries from natural language without reinforcement learning,'' \emph{CoRR}, 2017.

\bibitem{IRNet19}
J.~Guo, Z.~Zhan, Y.~Gao, Y.~Xiao, J.~Lou, T.~Liu, and D.~Zhang, ``Towards complex text-to-sql in cross-domain database with intermediate representation,'' in \emph{{ACL}}, A.~Korhonen, D.~R. Traum, and L.~M{\`{a}}rquez, Eds., 2019,  4524--4535.

\bibitem{Bogin19}
B.~Bogin, J.~Berant, and M.~Gardner, ``Representing schema structure with graph neural networks for text-to-sql parsing,'' in \emph{{ACL}}, 2019,  4560--4565.

\bibitem{BoginGB19}
B.~Bogin, M.~Gardner, and J.~Berant, ``Global reasoning over database structures for text-to-sql parsing,'' in \emph{{EMNLP-IJCNLP}}, 2019,  3657--3662.

\bibitem{RATSQL20}
B.~Wang, R.~Shin, X.~Liu, O.~Polozov, and M.~Richardson, ``{RAT-SQL:} relation-aware schema encoding and linking for text-to-sql parsers,'' in \emph{{ACL}}, 2020,  7567--7578.

\bibitem{GAP21}
P.~Shi, P.~Ng, Z.~Wang, H.~Zhu, A.~H. Li, J.~Wang, C.~N. dos Santos, and B.~Xiang, ``Learning contextual representations for semantic parsing with generation-augmented pre-training,'' in \emph{{AAAI}}, 2021,  13\,806--13\,814.

\bibitem{SmBop20}
O.~Rubin and J.~Berant, ``Smbop: Semi-autoregressive bottom-up semantic parsing,'' \emph{CoRR}, 2020.

\bibitem{LGESQL20}
R.~Cao, L.~Chen, Z.~Chen, Y.~Zhao, S.~Zhu, and K.~Yu, ``{LGESQL:} line graph enhanced text-to-sql model with mixed local and non-local relations,'' in \emph{{ACL/IJCNLP}}, 2021,  2541--2555.

\bibitem{resdsql23}
H.~Li, J.~Zhang, C.~Li, and H.~Chen, ``Resdsql: Decoupling schema linking and skeleton parsing for text-to-sql,'' in \emph{AAAI}, 2023.

\bibitem{GPT4}
OpenAI, ``{GPT-4} technical report,'' \emph{CoRR}, 2023.

\bibitem{DINSQL23}
M.~Pourreza and D.~Rafiei, ``{DIN-SQL:} decomposed in-context learning of text-to-sql with self-correction,'' \emph{CoRR}, vol. abs/2304.11015, 2023.

\bibitem{SQL-PaLM23}
R.~Sun, S.~{\"{O}}. Arik, H.~Nakhost, H.~Dai, R.~Sinha, P.~Yin, and T.~Pfister, ``Sql-palm: Improved large language model adaptation for text-to-sql,'' \emph{CoRR}, vol. abs/2306.00739, 2023.

\bibitem{Yu18}
T.~Yu, R.~Zhang, K.~Yang, M.~Yasunaga, D.~Wang, Z.~Li, J.~Ma, I.~Li, Q.~Yao, S.~Roman, Z.~Zhang, and D.~R. Radev, ``Spider: {A} large-scale human-labeled dataset for complex and cross-domain semantic parsing and text-to-sql task,'' in \emph{EMNLP}, 2018,  3911--3921.

\bibitem{topk18}
A.~Fan, M.~Lewis, and Y.~N. Dauphin, ``Hierarchical neural story generation,'' in \emph{{ACL}}, 2018,  889--898.

\bibitem{GimpelBDS13}
K.~Gimpel, D.~Batra, C.~Dyer, and G.~Shakhnarovich, ``A systematic exploration of diversity in machine translation,'' in \emph{{EMNLP}}, 2013,  1100--1111.

\bibitem{LiGBGD16}
J.~Li, M.~Galley, C.~Brockett, J.~Gao, and B.~Dolan, ``A diversity-promoting objective function for neural conversation models,'' in \emph{{NAACL}}, 2016,  110--119.

\bibitem{LiJ16}
J.~Li and D.~Jurafsky, ``Mutual information and diverse decoding improve neural machine translation,'' \emph{CoRR}, vol. abs/1601.00372, 2016.

\bibitem{SummaReranker22}
M.~Ravaut, S.~R. Joty, and N.~F. Chen, ``Summareranker: {A} multi-task mixture-of-experts re-ranking framework for abstractive summarization,'' in \emph{{ACL}}, 2022,  4504--4524.

\bibitem{PairReranker22}
D.~Jiang, B.~Y. Lin, and X.~Ren, ``Pairreranker: Pairwise reranking for natural language generation,'' \emph{CoRR}, vol. abs/2212.10555, 2022.

\bibitem{Joint23}
W.~Shen, Y.~Gong, Y.~Shen, S.~Wang, X.~Quan, N.~Duan, and W.~Chen, ``Joint generator-ranker learning for natural language generation,'' in \emph{{ACL}}, 2023,  7681--7699.

\bibitem{MIGA22}
Y.~Fu, W.~Ou, Z.~Yu, and Y.~Lin, ``{MIGA:} {A} unified multi-task generation framework for conversational text-to-sql,'' \emph{CoRR}, vol. abs/2212.09278, 2022.

\bibitem{GAR23}
Y.~Fan, Z.~He, T.~Ren, D.~Guo, C.~Lin, R.~Zhu, G.~Chen, Y.~Jing, K.~Zhang, and X.~Wang, ``Gar: A generate-and-rank approach for natural language to sql translation,'' in \emph{{ICDE}}, 2023.

\bibitem{GenSql23}
Y.~Fan, T.~Ren, Z.~He, X.~S. Wang, Y.~Zhang, and X.~Li, ``Gensql: {A} generative natural language interface to database systems,'' in \emph{{ICDE}}, 2023,  3603--3606.

\bibitem{Controllable23}
X.~Zheng, H.~Lin, X.~Han, and L.~Sun, ``Toward unified controllable text generation via regular expression instruction,'' \emph{CoRR}, 2023.

\bibitem{Controlled23}
M.~Kim, H.~Lee, K.~M. Yoo, J.~Park, H.~Lee, and K.~Jung, ``Critic-guided decoding for controlled text generation,'' in \emph{{ACL}}, 2023,  4598--4612.

\bibitem{Controllable22}
H.~Zhang, H.~Song, S.~Li, M.~Zhou, and D.~Song, ``A survey of controllable text generation using transformer-based pre-trained language models,'' \emph{CoRR}, 2022.

\bibitem{Gupta18}
N.~Gupta and M.~Lewis, ``Neural compositional denotational semantics for question answering,'' in \emph{EMNLP}, 2018,  2152--2161.

\bibitem{Talmor18}
A.~Talmor and J.~Berant, ``The web as a knowledge-base for answering complex questions,'' in \emph{{NAACL-HLT}}, 2018,  641--651.

\bibitem{decomposition19}
H.~Zhang, J.~Cai, J.~Xu, and J.~Wang, ``Complex question decomposition for semantic parsing,'' in \emph{{ACL}}, 2019,  4477--4486.

\bibitem{Multi-hop19}
S.~Min, V.~Zhong, L.~Zettlemoyer, and H.~Hajishirzi, ``Multi-hop reading comprehension through question decomposition and rescoring,'' in \emph{{ACL}}, 2019,  6097--6109.

\bibitem{Break20}
T.~Wolfson, M.~Geva, A.~Gupta, Y.~Goldberg, M.~Gardner, D.~Deutch, and J.~Berant, ``Break it down: {A} question understanding benchmark,'' \emph{Trans. Assoc. Comput. Linguistics},  183--198, 2020.

\bibitem{ScienceBenchmark23}
Y.~Zhang, J.~Deriu, G.~Katsogiannis{-}Meimarakis, C.~Kosten, G.~Koutrika, and K.~Stockinger, ``Sciencebenchmark: {A} complex real-world benchmark for evaluating natural language to {SQL} systems,'' \emph{CoRR}, 2023.

\bibitem{BRIDGE20}
X.~V. Lin, R.~Socher, and C.~Xiong, ``Bridging textual and tabular data for cross-domain text-to-sql semantic parsing,'' in \emph{{EMNLP}}, 2020,  4870--4888.

\bibitem{BahdanauCB14}
D.~Bahdanau, K.~Cho, and Y.~Bengio, ``Neural machine translation by jointly learning to align and translate,'' in \emph{{ICLR}}, 2015.

\bibitem{seq2seq14}
I.~Sutskever, O.~Vinyals, and Q.~V. Le, ``Sequence to sequence learning with neural networks,'' in \emph{NeurIPS}, 2014,  3104--3112.

\bibitem{EditSQL19}
R.~Zhang, T.~Yu, H.~Er, S.~Shim, E.~Xue, X.~V. Lin, T.~Shi, C.~Xiong, R.~Socher, and D.~R. Radev, ``Editing-based {SQL} query generation for cross-domain context-dependent questions,'' in \emph{{EMNLP-IJCNLP}}, 2019,  5337--5348.

\bibitem{NatSQL21}
Y.~Gan, X.~Chen, J.~Xie, M.~Purver, J.~R. Woodward, J.~H. Drake, and Q.~Zhang, ``Natural {SQL:} making {SQL} easier to infer from natural language specifications,'' in \emph{{EMNLP}}, 2021,  2030--2042.

\bibitem{nitarshan22}
N.~Rajkumar, R.~Li, and D.~Bahdanau, ``Evaluating the text-to-sql capabilities of large language models,'' \emph{CoRR}, vol. abs/2204.00498, 2022.

\bibitem{aiwei23}
A.~Liu, X.~Hu, L.~Wen, and P.~S. Yu, ``A comprehensive evaluation of chatgpt's zero-shot text-to-sql capability,'' \emph{CoRR}, vol. abs/2303.13547, 2023.

\bibitem{Interleaving23}
Z.~Gu, J.~Fan, N.~Tang, S.~Zhang, Y.~Zhang, Z.~Chen, L.~Cao, G.~Li, S.~Madden, and X.~Du, ``Interleaving pre-trained language models and large language models for zero-shot {NL2SQL} generation,'' \emph{CoRR}, vol. abs/2306.08891, 2023.

\bibitem{multi-bert19}
R.~F. Nogueira, W.~Yang, K.~Cho, and J.~Lin, ``Multi-stage document ranking with {BERT},'' \emph{CoRR}, 2019.

\bibitem{Gao21}
L.~Gao, Z.~Dai, and J.~Callan, ``Rethink training of {BERT} rerankers in multi-stage retrieval pipeline,'' in \emph{{ECIR}}, D.~Hiemstra, M.~Moens, J.~Mothe, R.~Perego, M.~Potthast, and F.~Sebastiani, Eds., 2021,  280--286.

\bibitem{Austin21}
J.~Austin, A.~Odena, M.~I. Nye, M.~Bosma, H.~Michalewski, D.~Dohan, E.~Jiang, C.~J. Cai, M.~Terry, Q.~V. Le, and C.~Sutton, ``Program synthesis with large language models,'' \emph{CoRR}, 2021.

\bibitem{Cobbe21}
K.~Cobbe, V.~Kosaraju, M.~Bavarian, J.~Hilton, R.~Nakano, C.~Hesse, and J.~Schulman, ``Training verifiers to solve math word problems,'' \emph{CoRR}, 2021.

\bibitem{AlphaCode22}
Y.~Li, D.~H. Choi, J.~Chung, N.~Kushman, J.~Schrittwieser, R.~Leblond, T.~Eccles, J.~Keeling, F.~Gimeno, A.~D. Lago, T.~Hubert, P.~Choy, C.~de~Masson~d'Autume, I.~Babuschkin, X.~Chen, P.~Huang, J.~Welbl, S.~Gowal, A.~Cherepanov, J.~Molloy, D.~J. Mankowitz, E.~S. Robson, P.~Kohli, N.~de~Freitas, K.~Kavukcuoglu, and O.~Vinyals, ``Competition-level code generation with alphacode,'' \emph{CoRR}, 2022.

\bibitem{GPT3}
T.~B. Brown, B.~Mann, N.~Ryder, M.~Subbiah, J.~Kaplan, P.~Dhariwal, A.~Neelakantan, P.~Shyam, G.~Sastry, A.~Askell, S.~Agarwal, A.~Herbert{-}Voss, G.~Krueger, T.~Henighan, R.~Child, A.~Ramesh, D.~M. Ziegler, J.~Wu, C.~Winter, C.~Hesse, M.~Chen, E.~Sigler, M.~Litwin, S.~Gray, B.~Chess, J.~Clark, C.~Berner, S.~McCandlish, A.~Radford, I.~Sutskever, and D.~Amodei, ``Language models are few-shot learners,'' in \emph{NeurIPS}, 2020.

\bibitem{LeeCT19}
K.~Lee, M.~Chang, and K.~Toutanova, ``Latent retrieval for weakly supervised open domain question answering,'' in \emph{{ACL}}, 2019,  6086--6096.

\bibitem{KarpukhinOMLWEC20}
V.~Karpukhin, B.~Oguz, S.~Min, P.~S.~H. Lewis, L.~Wu, S.~Edunov, D.~Chen, and W.~Yih, ``Dense passage retrieval for open-domain question answering,'' in \emph{{EMNLP}}, 2020,  6769--6781.

\bibitem{XiongXLTLBAO21}
L.~Xiong, C.~Xiong, Y.~Li, K.~Tang, J.~Liu, P.~N. Bennett, J.~Ahmed, and A.~Overwijk, ``Approximate nearest neighbor negative contrastive learning for dense text retrieval,'' in \emph{{ICLR}}, 2021.

\bibitem{BERT19}
J.~Devlin, M.~Chang, K.~Lee, and K.~Toutanova, ``{BERT:} pre-training of deep bidirectional transformers for language understanding,'' in \emph{{NAACL-HLT}}, 2019,  4171--4186.

\bibitem{LiuMZX0Z20}
C.~Liu, Z.~Mao, T.~Zhang, H.~Xie, B.~Wang, and Y.~Zhang, ``Graph structured network for image-text matching,'' in \emph{{CVPR}}, 2020,  10\,918--10\,927.

\bibitem{multigrained22}
Z.~Fan, Z.~Wei, Z.~Li, S.~Wang, H.~Shan, X.~Huang, and J.~Fan, ``Constructing phrase-level semantic labels to form multi-grained supervision for image-text retrieval,'' in \emph{{ICMR}}, 2022,  137--145.

\bibitem{listwise07}
Z.~Cao, T.~Qin, T.~Liu, M.~Tsai, and H.~Li, ``Learning to rank: from pairwise approach to listwise approach,'' in \emph{{ICML}}, 2007,  129--136.

\bibitem{SQL2NL10}
G.~Koutrika, A.~Simitsis, and Y.~E. Ioannidis, ``Explaining structured queries in natural language,'' in \emph{{ICDE}}, 2010,  333--344.

\bibitem{Iyer16}
S.~Iyer, I.~Konstas, A.~Cheung, and L.~Zettlemoyer, ``Summarizing source code using a neural attention model,'' in \emph{{ACL}}, 2016.

\bibitem{Xu18}
K.~Xu, L.~Wu, Z.~Wang, Y.~Feng, and V.~Sheinin, ``Sql-to-text generation with graph-to-sequence model,'' in \emph{EMNLP}, 2018,  931--936.

\bibitem{Pobrotyn21}
P.~Pobrotyn and R.~Bialobrzeski, ``Neuralndcg: Direct optimisation of a ranking metric via differentiable relaxation of sorting,'' \emph{CoRR}, 2021.

\bibitem{Kingma14}
D.~P. Kingma and J.~Ba, ``Adam: {A} method for stochastic optimization,'' in \emph{{ICLR}}, 2015.

\bibitem{RoBERTa19}
Y.~Liu, M.~Ott, N.~Goyal, J.~Du, M.~Joshi, D.~Chen, O.~Levy, M.~Lewis, L.~Zettlemoyer, and V.~Stoyanov, ``Roberta: {A} robustly optimized {BERT} pretraining approach,'' \emph{CoRR}, 2019.

\bibitem{MRR99}
D.~A. Hull, ``Xerox {TREC-8} question answering track report,'' in \emph{{TREC}}, 1999.

\bibitem{GEO96}
J.~M. Zelle and R.~J. Mooney, ``Learning to parse database queries using inductive logic programming,'' in \emph{{AAAI}}, 1996,  1050--1055.

\bibitem{precise08}
A.~Simitsis, G.~Koutrika, and Y.~E. Ioannidis, ``Pr{\'{e}}cis: from unstructured keywords as queries to structured databases as answers,'' \emph{{PVLDB}}, no.~1,  117--149, 2008.

\bibitem{Zettlemoyer05}
L.~S. Zettlemoyer and M.~Collins, ``Learning to map sentences to logical form: Structured classification with probabilistic categorial grammars,'' in \emph{{UAI}}, 2005,  658--666.

\bibitem{NaLR14}
F.~Li and H.~V. Jagadish, ``Constructing an interactive natural language interface for relational databases,'' \emph{{PVLDB}}, no.~1,  73--84, 2014.

\bibitem{ATHENA16}
D.~Saha, A.~Floratou, K.~Sankaranarayanan, U.~F. Minhas, A.~R. Mittal, and F.~{\"{O}}zcan, ``{ATHENA:} an ontology-driven system for natural language querying over relational data stores,'' \emph{{PVLDB}}, no.~12,  1209--1220, 2016.

\bibitem{TEMPLAR19}
C.~Baik, H.~V. Jagadish, and Y.~Li, ``Bridging the semantic gap with {SQL} query logs in natural language interfaces to databases,'' in \emph{{ICDE}}, 2019,  374--385.

\bibitem{Guo19}
J.~Guo, Z.~Zhan, Y.~Gao, Y.~Xiao, J.~Lou, T.~Liu, and D.~Zhang, ``Towards complex text-to-sql in cross-domain database with intermediate representation,'' in \emph{{ACL}}, 2019,  4524--4535.

\bibitem{Wang20}
B.~Wang, R.~Shin, X.~Liu, O.~Polozov, and M.~Richardson, ``{RAT-SQL:} relation-aware schema encoding and linking for text-to-sql parsers,'' in \emph{{ACL}}, 2020,  7567--7578.

\bibitem{Yu20}
T.~Yu, C.~Wu, X.~V. Lin, B.~Wang, Y.~C. Tan, X.~Yang, D.~R. Radev, R.~Socher, and C.~Xiong, ``Grappa: Grammar-augmented pre-training for table semantic parsing,'' \emph{CoRR}, 2020.

\bibitem{IKnowSQL23}
Y.~Fan, T.~Ren, D.~Guo, Z.~Zhao, Z.~He, X.~S. Wang, Y.~Wang, and T.~Sui, ``An integrated interactive framework for natural language to {SQL} translation,'' in \emph{{WISE}}, vol. 14306, 2023,  643--658.

\bibitem{CatSQL23}
H.~Fu, C.~Liu, B.~Wu, F.~Li, J.~Tan, and J.~Sun, ``Catsql: Towards real world natural language to {SQL} applications,'' \emph{Proc. {VLDB} Endow.}, vol.~16, no.~6,  1534--1547, 2023.

\end{thebibliography}

\end{document}